\documentclass[12pt, preprint]{aastex}
\usepackage{amsmath}
\usepackage{graphicx}
\usepackage{subfigure}

\def\cc{cm$^{-3}$}
\def\kms{km s$^{-1}$}
\def\s{s$^{-1}$}

\def\h2{H$_2$}
\def\n2h{N$_2$H$^+$}

\def\13co{$^{13}$CO}
\def\c18o{C$^{18}$O}

\def\lp{\>\> .}
\def\lc{\>\> ,}

\newcommand{\X}{\mathcal{X}}

\def\cm2{cm$^{-2}$}

\def\mic{$\mu$m}

\def\h2{H$_2$}

\def \no{n$_1$}
\def \n2{n$_2$}

\def \N2{N$_2$}

\def \p2{P$_2$}

\shorttitle{HI Narrow Self--Absorption}
\shortauthors{Li and Goldsmith}

\begin{document}
\title{HI Narrow Self--Absorption in Dark Clouds}

\author{D. Li\altaffilmark{1,2} and P. F. Goldsmith\altaffilmark{2}}

\altaffiltext{1}{Center for Astrophysics, 60 Garden Street, 
Cambridge MA 02138, dli@cfa.harvard.edu }
\altaffiltext{2}{National Astronomy and Ionosphere Center, Department of 
Astronomy, Cornell University,\\ Ithaca NY 14853}

\begin{abstract}
We have used the Arecibo telescope to carry out an survey
of 31 dark clouds in the Taurus/Perseus region for narrow absorption
features in HI ($\lambda$ 21cm) and OH (1667 and 1665 MHz) emission.  
We detected HI narrow self--absorption (HINSA) in 77$\%$ of 
the clouds that we observed. 
HINSA and OH emission, observed simultaneously are remarkably well 
correlated.
Spectrally, they have the same nonthermal line width and the same 
line centroid velocity. 
Spatially, they both peak at the optically--selected central
position of each cloud, and both fall off toward the cloud edges. 
Sources with clear HINSA feature have also been observed in transitions
of CO, \13co, \c18o, and CI. 
HINSA exhibits better correlation with molecular tracers than with CI.

The line width of the absorption feature, together with analyses
of the relevant radiative transfer provide upper limits to the 
kinetic temperature of the gas producing the HINSA. 
Some sources must have a temperature close to or lower than 10 K. 
The correlation of column densities and line widths of HINSA 
with those characteristics of
molecular tracers suggest that a significant fraction of the atomic 
hydrogen is located in the cold, well--shielded portions of molecular
clouds, and is mixed with the molecular gas.

The average number density ratio  [HI]/[\h2] is $1.5\times10^{-3}$.
The inferred HI density appears consistent with but is slightly higher
than the value expected in steady state equilibrium between formation of
HI via cosmic ray destruction of H$_2$ and destruction via formation
of H$_2$ on grain surfaces.
The distribution and abundance of atomic hydrogen
in molecular clouds is a critical test of dark cloud chemistry 
and structure, including the issues of grain surface reaction rates, PDRs, 
circulation, and turbulent diffusion.

\end{abstract}

\setcounter{footnote}{0}
%
%

\section{INTRODUCTION}
	
Two relatively distinct phases are generally assumed to exist in the 
neutral interstellar medium (ISM): atomic and molecular. 
The atomic phase of the ISM, consisting mainly of hydrogen atoms, 
is traced by the HI hyperfine transition at $\lambda$ 21cm. 
The molecular phase of the ISM, whose major component -- molecular hydrogen
-- lacks a permanent electric dipole moment and readily excited transitions
at temperatures generally encountered, is primarily traced by emission 
from rarer molecular species such as carbon monoxide. 
The  conversion from atomic to molecular forms occurs on dust grains, 
where atomic hydrogen sticks and forms \h2. 
This exothermic reaction releases \h2 into the gas and keeps molecular 
clouds molecular.
Dissociative processes maintain a population of atoms even inside molecular clouds.
Recently, there has been increased interest in atomic species, which
prove to be important probes. 
Two important examples are CI, which is accessible at submillimeter 
wavelengths \citep*[e.g.\ ][]{phillips80,huang99,plume00}, and OI, which can be observed 
in the far infrared \citep*[e.g.\ ][]{herr97, kram98, lise99, lis01}.  

Inside molecular clouds, dissociating UV photons are blocked both by 
grains and by \h2 line absorption \citep{hws71}. 
A significant HI population exists inside molecular clouds
maintained by cosmic ray destruction of \h2 and additionally as a
remnant of the \h2 formation process in a chemically young cloud. 
The atomic hydrogen component inside molecular clouds has fractional 
abundance ([H]/[\h2]) of $\simeq$ 0.1$\%$ (discussed in Section \ref{nhih2}) 
and is thus the third most abundant gas phase species, after \h2 and He.
Because the balance of HI and \h2 involves grain surface reactions,
the density of atomic HI in dark clouds serves as a test of
complete chemical networks with reactions both in the gas phase and 
on grain surfaces.
The HI abundance in well--shielded regions can be increased by
relatively rapid turbulent diffusion \citep{willacy02}, or by
general mass circulation \citep{chie89}.
It is therefore important to establish the presence of HI in the molecular
ISM and to determine accurately its abundance. 

\begin{figure}                                 
\plotone{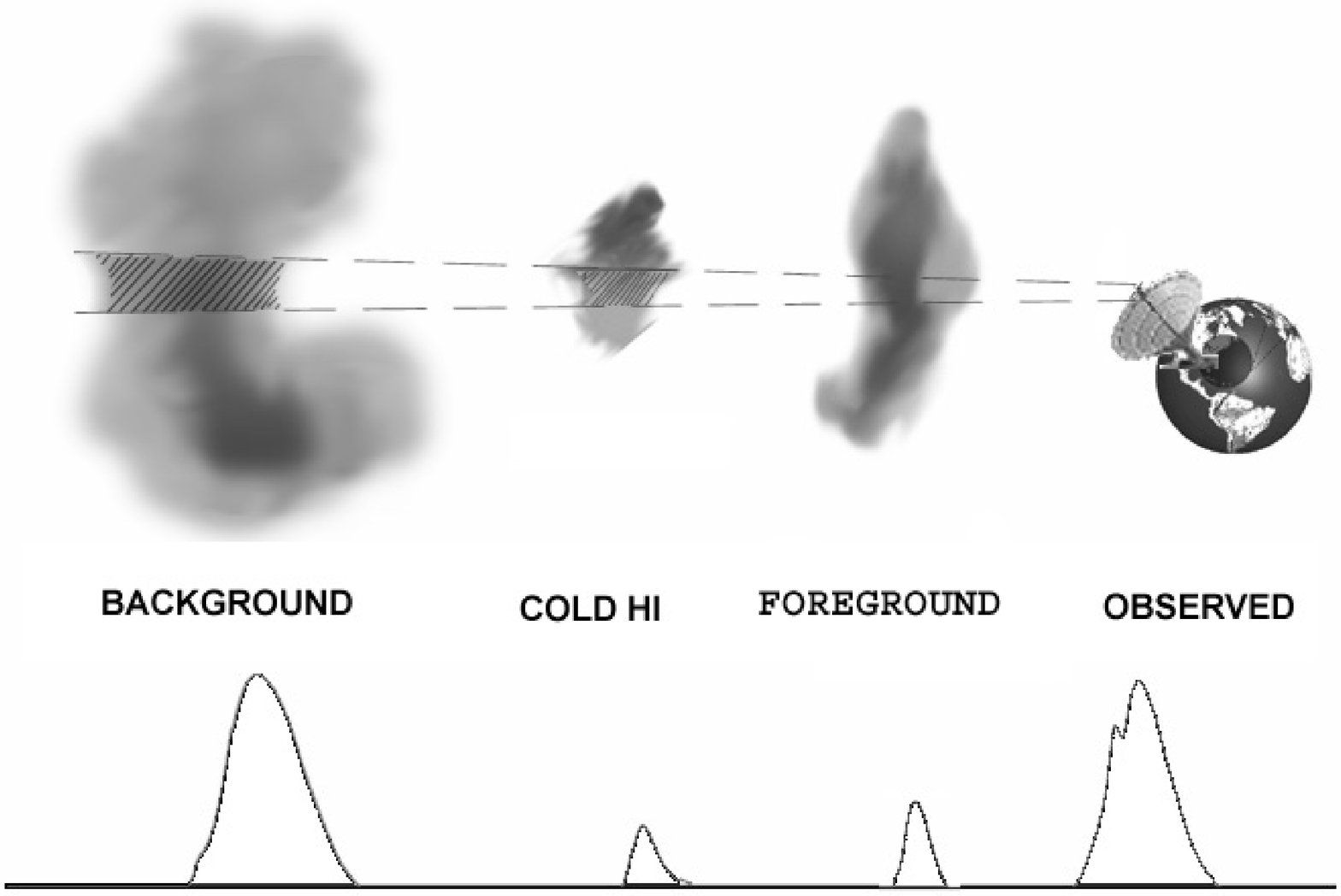}
\caption{Three--body radiative transfer configuration.}
\label{fig:abs}
\end{figure}

A unique probe of this component is HI narrow (which we define to be
less than that of CO) line width self--absorption, which we denote HINSA.
The absorption dips seen in the spectra of $\lambda$ 21cm emission 
are often denoted HI 
self--absorption. 
The prefix `self' is widely used to differentiate this phenomenon
from absorption against a background continuum source. 
For the origins of HI absorption toward nearby dark clouds,
\emph{separate} galactic background 
HI emission and cold foreground HI material are both needed.
A typical configuration is  shown in Figure~\ref{fig:abs}. 

The existence of cold HI associated with dark clouds was recognized
more than 25 years ago. 
\citet{knapp74} conducted a survey of 88 dark clouds and detected 
absorption features in fewer than half of them. 
The optical depth of cold HI was derived from the profiles of the 
absorption lines. 
The molecular content of the clouds observed was traced by their dust 
extinction.
The HI fractional abundance, [H]/[\h2] was thus determined to be 
1\% to 5\%.
In terms of the observational limits and conclusions, this early work 
typifies most self--absorption studies that have followed.

With the 140ft (43m) telescope, Knapp's survey has an angular resolution
of 21\arcmin\ and a velocity resolution of $\sim$0.5 \kms.
Similar resolutions have been achieved with the 85ft (26m) antenna
at Hat Creek \citep{goodman94} and the 120ft (37 m) antenna of 
the Haystack Observatory (Myers et al. 1978). 
The 76m Lovell telescope (12\arcmin\ angular resolution, $\sim$0.5 \kms\ 
velocity resolution; the same convention in what follows) has been 
employed for a study of six dark clouds in the Lynds (1962)
catalogue (McCutcheon, Shuter \& Booth  1978) and the 
Riegel--Crutcher cloud (Montgomery, Bates \& Davies 1995). 
Additional studies have been conducted with the Effelsberg 100m telescope 
(9\arcmin, 0.5 \kms). 
The Taurus molecular cloud TMC1 has been mapped by Wilson and Minn (1977). 
The complex region around B18, also known as Kutner's cloud, 
has been mapped by Batrla, Wilson \& Rache (1981)
and by P\"{o}ppel, Rohlfs \& Celnik (1983). 
The Arecibo telescope with the line feed system (3\arcmin, 1 \kms) 
has been used to study HI absorption  \citep*[e.g.\ ][]{burton78,
baker79, bania84}. Baker \& Burton (1979)
also raise the possibility that the absorption can help 
resolve the near--far ambiguity in kinematic distances, a topic which 
is further explored by Jackson et al.\ (2002).

\begin{figure}                                 
\epsscale{0.8}
\plotone{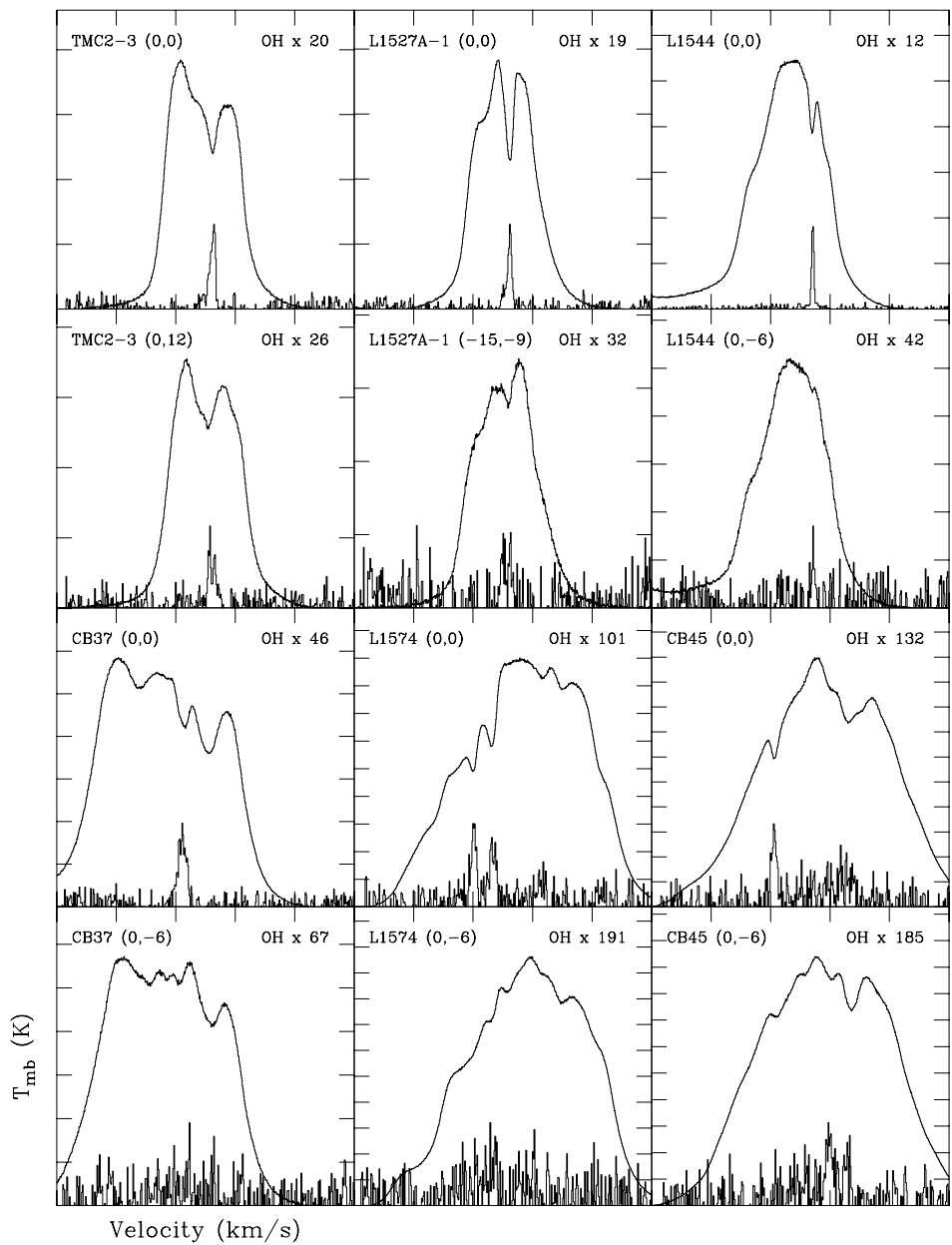}
\caption{Selected sources together with associated OFF positions. 
The narrow HI absorption feature and OH 1667 MHz emission line
are shown.
The X axes are centered on the $V_{lsr}$ of the HINSA (given in Table 2)
and each tick mark represents 10 \kms. 
The units of the Y axes are main beam antenna temperature
and each tick mark represents 10 K.
The number following ``OH x'' is the factor by which the OH line has been 
multiplied in order to be visible on the plot.
}
\label{dipoff}
\end{figure}

Interferometers have also been used to obtain higher angular resolution, 
but the usual penalty has been lower velocity resolution to augment the 
sensitivity. 
Van der Werf et al.\ mapped L134 (1988) and L1551 (1989) with the
DRAO and the VLA (1.5\arcmin, 1.3 \kms). 
The interferometer studies are well--suited for mapping structures 
having a scale of arc--minutes. 
But their velocity resolutions of about 1.5 \kms\ can easily miss
or suppress narrow absorption features, which occupy at most
a couple of velocity channels in such spectra.
To compound this problem, the HI emission is structured.
Multiple peaks and wide troughs are not rare in galactic 21c profiles 
(see, e.g. Figure~\ref{dipoff}). 
If a map is based on the integrated area of an absorption line (or lines), 
the wide troughs which may not be associated with dark clouds 
will be more prominent than are the narrow features. 
Reliable analysis of HINSA profile requires a velocity resolution better 
than 0.3 \kms.

The general scientific objectives of narrow line HI absorption studies are
to explain its origin and to determine the abundance of atomic hydrogen.
For the first question, the association of HINSA
with dark clouds is inconclusive in the literature. 
On the positive side, Sherwood \& Wilson (1981) find a good correlation 
with extinction in TMC1.  
McCutcheon et al.\ (1978) give a higher detection rate (6--8 out 
of 11 Lynds clouds) than Knapp (1974). 
Cappa de Nicolau \& Poppel (1991) find HINSA in the `darkest' cores in the 
CrA complex embedded in a HI emission ridge. 
On the other hand, the association between HI  absorption features 
and dense regions is ambiguous in a recent DRAO HI survey, 
in which Gibson et al.\ (2000) find 
cloud--like absorption structures both correlated and uncorrelated with
CO emission. 
They label these features HISA, shorthand for HI self--absorption. 
This is the situation in which the distinction between HISA and HINSA 
must be made. 
As seen in the DRAO survey, the HISA  probably reflects
temperature fluctuations in the atomic ISM. 
The HISA is spectrally wider and flatter and could have its origin in the 
same location as the 21cm emitting gas, making its name appropriate. 
HINSA refers to narrower absorption features, which are produced by cold 
foreground molecular clouds and which are not prominent in the DRAO survey 
with 1.3 \kms\ velocity resolution. To further complicate the
issue, HI has also been seen in {\em emission} in possible halos
around molecular clouds, such as B5 \citep{wannier91}. 
HI halos are a distinct environment for atomic hydrogen compared to 
either HISA or HINSA, and they are easily distinguished observationally 
(see Section~\ref{discussion}).

The difficulties in obtaining the HI abundance through HINSA are 
three--fold. 
First, the early studies are hampered mainly by limited knowledge of 
the dark cloud itself. 
The  \h2 column density is often obtained from low angular resolution 
extinction data (e.g.\ Knapp 1974, Batrla et al.\ 1981) or
H$_2$CO (e.g. Wilson \& Minn 1977, Poppel et al.\ 1983), a molecule
whose fractional abundance is not very certain.
Second, with one profile, it is impossible to obtain both the 
optical depth and the spin temperature accurately. 
An assumption about the spin temperature is often made, which may not 
be realistic. 
Third, the issue of foreground emission is often ignored. 

With some or all of the three uncertainties, the [H]/[\h2] 
ratio has been derived to be ranging from a 
few percent (e.g.\ Knapp 1974; Saito, Ohtani \& Tomita 1978), 
to 5$\times$10$^{-4}$ \citep{winnberg80}.

The requirements of reasonable angular resolution, good sensitivity, and 
high frequency resolution make the upgraded Arecibo Gregorian system 
a valuable tool with which to study the HINSA. 
The large instantaneous bandwidth allows OH 1665 and 1667 MHz spectra
to be obtained simultaneously with that of HI.
%
%
We describe the new observations including a HI survey to examine the 
correlation between HINSA and OH in dark clouds and complementary mapping 
in carbon monoxide isotopologues, OH, and CI in Section 2. 
In Section 3 we present a three--component model for radiative transfer 
and correction for foreground material, allowing accurate HI column 
densities to be obtained.
We analyze OH and \c18o emission in Section 4, and the results of 
our survey and observations of L1544 in Section 5.
We review the general issue of atomic hydrogen in molecular clouds
in Section 6, discuss our results in Section 7, and summarize our
conclusions in Section 8.

%
%
%
%
\section{Observations 
\label{obs}}
	
The sources included in our survey of HI in dark cloud cores
are mainly chosen from a dark cloud catalogue based solely on  
optical obscuration \citep{lee99}. 
The observed cores meet the following constraints: 
(1) Right Ascension 2h -- 6h, 
(2) angular diameter close to or larger than 3\arcmin, 
(3) declination angle 0\arcdeg\ -- 35\arcdeg, 
(4) other observing considerations, such as minimizing slew time.
Twenty eight cores have thus been chosen for this Arecibo survey of dark 
clouds
Their names, coordinates, presence of HINSA, and association with a
young stellar object are given in Table 1. 
In addition, two well studied dark clouds in Perseus, 
Barnard 1 (B1) and Barnard 5 (B5), are included in our survey. 
B1 is one of the few dark clouds with a positive Zeeman detection 
\citep{goodman89}. 
The physical conditions of B5 have been well determined through mapping of 
CO and its isotopologues \citep{young82}. 
The inclusion of these two sources also extends the length of the nightly
observing session at Arecibo, due to their early rise times.
Because there is no a priori knowledge regarding the HINSA properties of
B1 and B5, they should not bias our study of the correlation between 
HINSA and dark clouds.
TMC1 is a well studied region with both strong OH emission and HI absorption.
We observe TMC1CP, the cyanopolyyne peak of TMC1 \citep{pratap97},
as a secondary calibrator source to check the consistency of
data taken at different times. 
Because it is a previously known strong HINSA source (Wilson \& Minn 1977), 
TMC1CP is excluded from source statistics presented later. 
In summary, our sample has 30 sources plus TMC1CP as a calibrator.

L--band observations of nearby dark clouds were conducted at Arecibo
in September 1999, February 2001 and January 2002. 
The 3\arcmin\ beam size corresponds to about 0.13 pc at the 140 pc distance
of the Taurus dark cloud complex.
Most individual nearby clouds fill the main beam at this frequency. 
The Arecibo spectral line correlator is configured to 
observe HI and OH (1665 MHz and 1667 MHz) simultaneously. 
The two lines of OH and the 21cm line of HI are recorded with a channel 
width of 0.382 kHz, which after smoothing yields a velocity resolution of 
0.14 \kms\ for OH and 0.16 \kms\ for HI. 
The remaining section of the correlator is employed to record the HI 
profile at a lower resolution with a wider bandwidth. 
This is necessary for obtaining a good spectral baseline when the 
HI emission toward some sources is nearly as wide as the bandpass at 
the higher resolution, as does occur for some sources (e.g. L1578--2).             

The data were taken in a total power (ON scan only) observing mode.
The prevalence of background HI emission and its variation on an arc--minute 
scale make finding clean OFF positions very questionable. 
Especially in regions of extended emission, such as the Taurus dark cloud 
complex, the use of an OFF position can lead to large uncertainties 
caused by variation in the emission. Sources such as CB 37, shown in 
Figure~\ref{dipoff}, illustrate such fluctuations across the band with 
amplitude comparable to the narrow absorption feature. 
In a study of HI envelopes, Wannier et al.\ (1983) find in their 
sample that the average variation of HI emission across a
cloud boundary is 10.9 K at the molecular emission velocity.
Fortunately, the narrowness of the absorption dip of interest, 
together with the stability of the Arecibo system, makes possible 
the measurement of the line profile even without an OFF scan 
(see Section 3). 
In addition to the reduction in the uncertainty due to emission in
the OFF scan, the stability of the system also allows a factor of 2
reduction in the statistical fluctuations by eliminating any switching.
	
The correction for the elevation dependence of the antenna gain and 
the absolute calibration are carried out by observations of quasars.
By cross--scanning these strong point sources, beam maps are
also obtained to measure the size of the main beam.
In all the data reduction which follows, we treat the slightly
elliptical beam to be a circular one with the same solid angle.
The derived aperture efficiency is 64\% and the main beam efficiency 60\%.
All L--band data are corrected for the main beam efficiency. 
This is exact if the source just fills the main beam, and 
appears to be close to an optimum strategy
for a one--parameter calculation.

Each ON source position was observed for an integration time of 5 to 10
minutes. 
At the current sensitivity of Arecibo ($T_{sys} \sim 35$ K), Galactic 
HI profiles without ambiguity caused by noise can be obtained in 
an even shorter time. 
The RMS noise level of less than 0.1 K in all our spectra is set by 
requirement of detecting OH lines, which typically have antenna 
temperatures in the range of 0.5 K to 1 K.
The beam efficiency for data taken in Feb 2001 is significantly higher than 
it was previously, as a result of surface readjustment for the primary.
Relative calibration and consistency checks are carried out 
using the strongest source, TMC1CP, which was observed both before and 
after the surface readjustment. 
The data from 2001 observations were scaled to those obtained at the earlier 
epoch, when extensive quasar calibration observations were made.

To determine the \h2 column density reliably, \c18o\ data on cores
with clear HINSA were obtained in October 2000 at the Five College Radio 
Observatory (FCRAO). 
The `footprint' size of the 16 element SEQUOIA focal plane array is 
5.9\arcmin$\times$5.9\arcmin. We use four pointings to make a beam sampled
map of each source and convolve it with the Arecibo beam. 
The data are corrected for the main beam efficiency to give the best 
estimate of the total mass in the beam as discussed above. 
By combining \c18o\ and HI data, the abundance [H]/[\h2] can be obtained. 
At FCRAO, CO and \13co spectra were also obtained for comparison with HI 
absorption line profiles.

The 492 GHz ground state fine--structure line of atomic carbon (CI)
has also been observed by the Submillimeter Wave Astronomy Satellite (SWAS) 
toward clouds with clear HINSA features.
In PDR models, CI should be abundant only in low extinction regions.
We compare its line characteristics to those of HINSA to examine the 
importance of photodissociation in maintaining populations of low 
temperature HI.
%
%
%
%
\section{Three--component Radiative Transfer and Foreground Correction} 
\label{rad}
		
\subsection{Radiative Transfer}
The standard approach for analyzing absorption is to reconstruct the 
emission spectrum from observing one or multiple `OFF' positions, 
where the absorption is absent.
As noted in section~\ref{obs}, identifying such OFF positions is problematic 
for HINSA. 
However, the emission spectrum can still be estimated for HINSA because 
the absorption line is much narrower (typically a factor of 20) than the 
background emission. The instrumental baseline is both flat and stable,
which allow us to use only the ON spectra. 

The reconstruction is carried out by fitting a 5th--order 
polynomial across the
frequencies affected by the absorption in the HINSA spectrum to define 
the profile as if there were no absorption present. The details
and uncertainty associated with such a fit will be discussed in the
following subsection.
 

The classical problem of deriving two quantities, the optical depth and
the excitation temperature from one profile is compounded by the 
possible foreground contamination, which adds more unknowns to the mix.
A careful look at the radiative transfer is necessary before any 
quantitative analysis of the HINSA can be carried out with confidence.

A three--component model is outlined in Figure~\ref{fig:abs}. 
It represents a general view of the HI absorption in the galaxy. 
The parameters are defined as
\begin{itemize}
\item $T_b$:  	background HI temperature , 
\item $T_f$:  	foreground HI temperature,
\item $\tau_b$:	background HI optical depth ,
\item $\tau_f$:	foreground HI optical depth,
\item $T_x $:	excitation temperature of HI in the dark cloud , 
\item $\tau$: 	optical  depth of HI in the dark cloud, 
\item $T_c $: 	continuum temperature, including the cosmic background and 
		Galactic continuum emission. 
\end{itemize}

The antenna temperature produced by the three--component system is
\begin{equation}
%
%
T_A=[T_c e^{-\tau_b} + T_b(1-e^{-\tau_b})]e^{-\tau}e^{-\tau_f}+T_x(1-e^{-\tau})e^{-\tau_f}+T_f(1-e^{-\tau_f})\lp
\end{equation}

To focus on the effect of the intervening dark cloud,
the
Galactic atomic gas is assumed to be of uniform temperature ($T_h$)
and small optical depth ($\tau_h$) at 21 cm. 
We can then write
\begin{equation}
%
%
\label{th}
T_h=T_b=T_f \lc
\end{equation}
and
\begin{equation}
%
%
\label{tauh}
\tau_h=\tau_f+\tau_b \lp
\end{equation}
To consolidate the variables, we define 
\begin{equation}
%
%
\tau_b=p\tau_h \lp
\end{equation}

For our ON spectra, a linear 
baseline is removed along with the passband shape.
Since the background continuum is flat across the band in our observations, it is 
also removed through our baseline fitting procedure. 
Thus,
the effective spectrum is given by
\begin{equation}
T_R = T_A-T_c~.
%
%
\end{equation}
%

Another quantity is obtained from the spectrum by fitting a
polynomial to the portion of the spectrum without the absorption.
We call this quantity $T_{HI}$,
\begin{equation}
%
%
T_{HI}=(T_h-T_c)(1-e^{-\tau_h}) \lc
\label{thi}
\end{equation}
 which is the HI 
temperature that would be observed if there were no absorbing cold cloud
and under the assumption outlined in equations~\eqref{th} and \eqref{tauh}.

Taking the difference of $T_{HI}$ and $T_R$, we can obtain, as a
positive quantity, the absorption temperature
\begin{eqnarray}
%
%
\nonumber T_{ab} &=&T_{HI}-T_R  \\
	 &=& [(T_c-T_h)e^{-\tau_h} + (T_h-T_x)e^{-\tau_f}](1-e^{-\tau}) \lp
\label{tabraw}
\end{eqnarray}
For small foreground and background HI opacities and with the definitions of $p$ 
and $T_{HI}$, the absorption temperature can be written as
\begin{equation}
%
%
%
T_{ab}= [pT_{HI} + (T_c-T_x)(1 - \tau_f)](1-e^{-\tau})  \lc
\label{tab}
\end{equation}
where $T_{ab}$ and $T_{HI}$ are observables and $\tau$ is
the quantity desired from the analysis. 
The explicit dependence
on the HI emission temperature $T_h$ is eliminated.
Let us examine the
remaining unknowns, $p$, $T_c$, $T_x$, and $\tau_f$.

The fraction of the HI along the line of sight lying beyond the cloud
responsible for the absorption, $p$,
can be calculated from a model of the 
local HI distribution. 
To first order, the Galactic HI disk can be approximated by a single 
Gaussian with a full width half maximum (FWHM) vertical extent $z$ equal 
to 360 pc \citep{lock84}. 
The percentage of HI in the background is thus given by the 
complementary error function
\begin{equation}
%
%
p = erfc[\sqrt{4\ln(2)}D\sin(b)/z] \lc
\label{p}
\end{equation}
where D is the distance to the absorbing cloud, b is its galactic latitude,
and $erfc(x)= 1 - \frac{2}{\sqrt{\pi}} \int_0^x e^{-t^2} dt$.

The Galactic background emission is estimated to be about 0.8 K by 
extrapolating the standard interstellar radiation field (ISRF) to L--band 
\citep*[e.g.\ ][]{winnberg80}. 
$T_c$ = 3.5 K is thus used in the analysis which follows.

The excitation temperature of the 21cm line in dark clouds, $T_x$,
is determined by the balance between collisional and radiative processes.  
Ignoring radiative trapping, the solution to the two level problem 
is given by \cite{purcell56} as
\begin{equation}
%
%
T_x = \frac{y}{1+y}T_k + \frac{1}{1+y}T_c \lc
\end{equation}
where $T_x$, the excitation temperature of the 21cm line in
our terminology, 
is also called the spin temperature, $T_k$ is the kinetic temperature of 
the gas in the dark cloud, and $T_c$ is the effective temperature of the
radiation field given above.
The quantity y is defined by
\begin{equation}
%
%
y = \frac{h\nu}{kT_k}\frac{C_{ul}}{A_{ul}} \lc
\end{equation}
where $C_{ul}$ is the collisional deexcitation rate and $A_{ul}$ is the 
spontaneous decay rate, $2.85\times10^{-15}$ \s\ 
\citep{wild52}.
There have been various calculations of the collision rates for the
deexcitation of the 21cm line, but the most relevant are the spin--exchange
collisions with another hydrogen atom \citep{field58, all69}.
The collisional deexcitation rates calculated with a full quantum scattering
code in the latter reference are considerably smaller than values 
determined earlier, and for a temperature $T_k$ = 10 K yield a value 
y = 5.5$n_1$, where $n_1$ is the atomic hydrogen density in cm$^{-3}$.  

For y $>>$ 1, the excitation temperature approaches the kinetic temperature
of the gas.  
This requires an atomic hydrogen density of at least 1 \cc.
As we show in the following, this is a reasonable value for the atomic
hydrogen density.
Inside dark clouds, most of the hydrogen is in the form of molecules. 
But as discussed in Section \ref{nhih2}, a steady state HI density 
is maintained by cosmic ray destruction 
of \h2; this density is independent of the density of \h2. 
The standard model gives $n_1 \sim 1$ \cc, in reasonable 
agreement with our observations.
Thus, the assumption of thermalization is justified.
For lower densities, since $T_c$ is less than $T_k$,
the excitation temperature will be below the kinetic temperature. 

The effect of the foreground comes in two terms in equation~\eqref{tab},
and the change in $T_{ab}$ due to the foreground absorption is
equal to $(p-1)T_{HI}+(T_x-T_c)\tau_f$. The excitation temperature is close
to the kinetic temperature of dark clouds, which is always lower than
the HI emission temperature $T_{HI}$.
Since p is generally in the range 0.8 to 0.9 and since $\tau_f$ is only a few
tenths, $(1-p)T_{HI}$ is the dominant term for our HINSA observations.
This means that $T_{ab}$ becomes smaller due to
the existence of foreground HI gas. 
The uncertainty in the derived HINSA column density resulting from
our not knowing the exact value of $\tau_f$
will be discussed in the following subsection.

Finally, the peak optical depth of the absorption feature
at the 
centroid velocity
can be obtained using equation~\eqref{tab}
\begin{equation}
%
%
\label{tauab}
\tau_0 = \ln [ \frac{pT_{HI}+(T_c-T_x)(1-\tau_f)}{pT_{HI}+(T_c-T_x)(1-\tau_f)-T_{ab}} ] \lc
%
\end{equation}
which in turn gives column density of HINSA as
\begin{equation}
%
%
\frac{N(\text{HINSA})}{\text{cm}^{-2}} = 1.95\times10^{18} \tau_0 \frac{\Delta V}{\text{km s}^{-1}} \frac{T_k}{\text{K}}  \lc
\label{nhi}
\end{equation}
where $\Delta V$ is the FWHM 
of the absorption line from a Gaussian fit. 

In summary, the recipe we have used for obtaining the HINSA column 
density involves, first, fitting the emission line profile to determine
T$_{HI}$, second, using a model of the Galactic HI to determine p,
and
third, deriving the peak optical depth of the cold HI through 
equation~\eqref{tauab}. 
Fourth, we use equation~\eqref{nhi} with the line width fitted 
to the absorption feature, to determine the column density. 
By fitting the emission and absorption portions of the spectra
separately and using these two fits, the optical depth of 
the absorption alone can be derived without explicitly knowing
the background and foreground temperatures, as 
long as they are equal and the galactic HI emission is optically thin.

\subsection{Uncertainties \label{unc}}

The OH intensities are 10 to 200 times smaller than of HI emission.
At the current level of integration, which is chosen to ensure good OH detections,
the noise in HI spectra is insignificant. 
The uncertainties in the derived HINSA
column density result primarily from 
three aspects of the assumptions behind and modeling
of the radiative transfer.

First, the imperfect knowledge of the Galactic HI disk and 
the cloud distance results in  uncertainties in $p$.
For 5 clouds, the distance is not known and $p$ cannot be modeled.
The remaining clouds have an average $<p>=0.83$. 
The difference between unity and this average provides us
an idea of the uncertainty in $p$. 
According to equation~\eqref{tauab},
this results in a 30$\%$ uncertainty 
for the derived optical depth
and, in turn, the column density.

\begin{figure}[htp]                                 
 \centering 
     \subfigure{\includegraphics[width=.60\textwidth, angle=90]{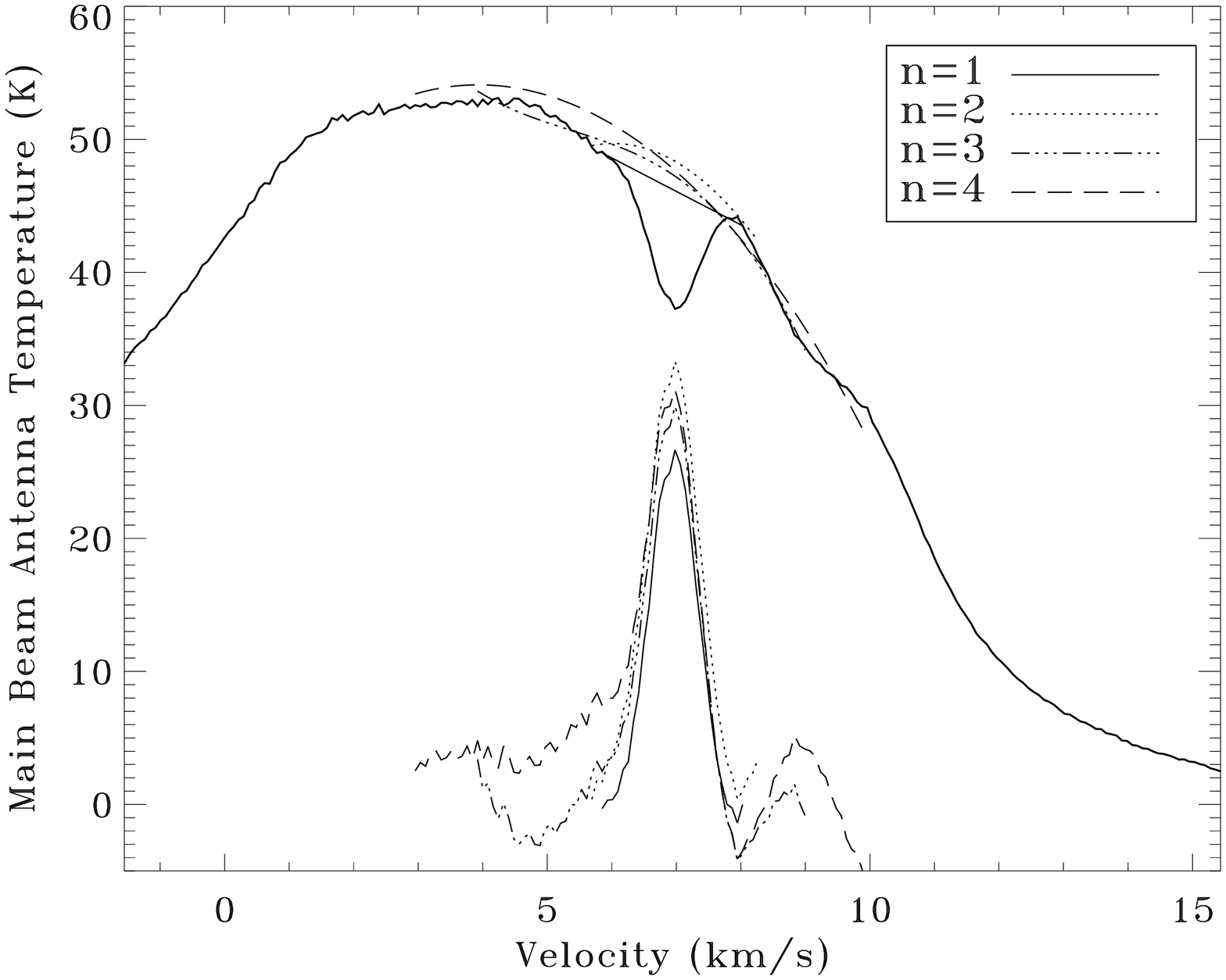}}\\
     \subfigure{\includegraphics[width=.60\textwidth, angle=90]{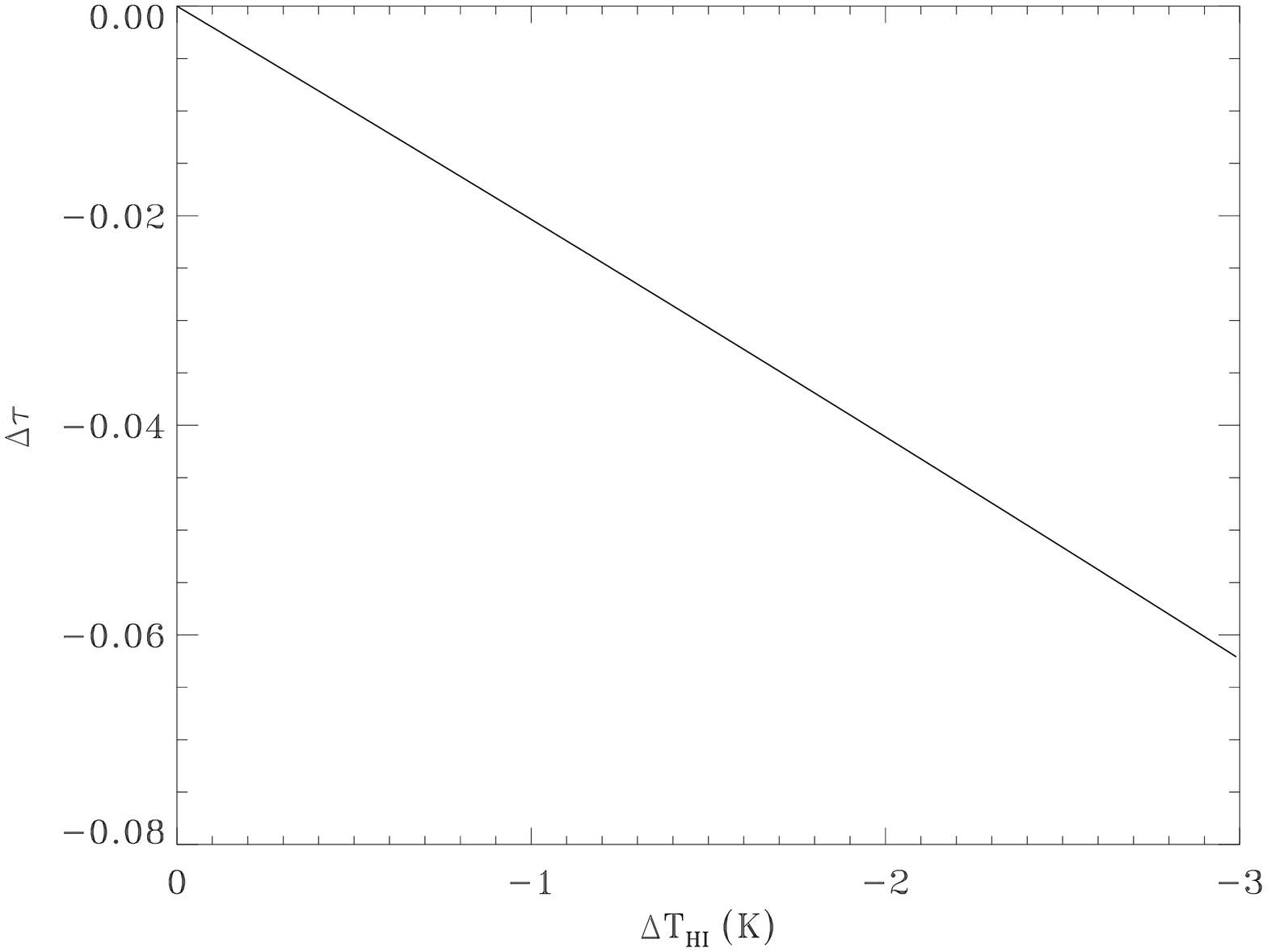}}
\caption{Upper panel: Fits to the background HI emission based on polynominal
of the $n$th order. 
The difference between the fitted HI emission for each polynomial order 
and the original spectrum (shown as an emission feature and enlarged by a 
factor of two) is also included. 
As discussed in the text, the polynomial order employed makes a 
modest difference in the line intensity derived.
Lower panel: the uncertainties in derived 
optical depth of HINSA ($\Delta \tau$) caused by
the fitting uncertainties of background HI emission ($\Delta T_{HI}$).}
\label{fig:rta}
\end{figure}

Second, the fitting to the background is somewhat arbitrary.
The upper panel of figure~\ref{fig:rta} demonstrates the effect of 
using different orders
of polynomial. 
The maximum difference in the fitted temperatures is 1.1 K and 
that in linewidth is 0.14 \kms. 
Higher orders of polynomial usually do not work well. 
A Gaussian fit to the background
is not feasible for sources with
very structured HI emission profiles, such as L1574
and L1633. 
The magnitude of the uncertainties also depends on the position
of absorption relative to the peak emission. 
When absorption occurs
closer to the emission center than the situation
shown in Figure~\ref{fig:rta}, the uncertainty in the antenna temperature 
becomes larger and the uncertainty in the linewidth becomes smaller.
We take 2 K as a representative value of uncertainty in the temperature and
0.15 \kms\ as that for the line width. 
These values are also included in Tables 2 and 3.
In the lower panel of figure~\ref{fig:rta} shows the change in the peak
HINSA optical depth produced by the uncertainty in the fitting of
the background HI emission. 
A deviation of 2 K in temperatures (both $T_{HI}$ and $T_{ab}$)
results in a change in $\tau_0$ by 0.04. This is 14$\%$ relative
to the average optical depth of the absorption.

Third, we consider the uncertainties in the $(T_c-T_x)(1-\tau_f)$ term.
In order to conform to the general assumptions of an optically thin HI emission
($\tau_h<1$), the value of $\tau_f$ is thus between 0 and about 0.1.
With $T_x=10$ K, $T_c=$ 3.5 K and typical numbers for $T_{HI}$, this corresponds 
to a change in $\tau$ of absorption of about 0.01, which is 4$\%$ relative to the
average optical depth.
We conclude that the uncertainty in the derived HINSA column density is dominated 
by systematic effects, and the overall uncertainty is approximately 50$\%$.

%
%
\section{Analysis Techniques for Spectral Lines with Small Opacities}
%
%
\subsection{OH Column Density}

From the definition of opacity, 
the column density of OH can be written as
\begin{equation}
%
%
N_{OH}=\frac{8 \pi k T_x \nu^{2}}{A_{1667} c^{3} h} \frac{16}{5} 
		\int \tau_{1667} dV \lc
\end{equation}
where $A_{1667}=7.778 \times 10^{-11}$ \s\ is the A coefficient of 
this transition and $T_x$ is its excitation temperature.
Under the assumptions of negligible optical depth and no background
emission, the column density can be calculated from the integrated intensity
(in K \kms) of the 1667 MHz line through \citep*[e.g.\ ][]{knapp73, turner71}
\begin{equation}
%
%
\label{noh}
\frac{N_{OH}}{\text{cm}^{-2}} = 2.22\times10^{14}\frac{\int T_{mb} (V) dV}{\text{K km s$^{-1}$}} \lp
\end{equation}	

The satellite component of this $\Lambda$ doubling line at 1665 MHz can 
be useful in determining the the optical depth and excitation temperature.
For emission without anomalies, the antenna temperatures of the two
components satisfy
\begin{equation}
%
%
\frac{T^{1667}_A}{T^{1665}_A} = \frac{ 1 - e^{\tau^{1667}}}
		{ 1 - e^{\tau_{1665}}}= \frac{ 1 - \X}{ 1 - \X^{0.552}} \lc
\label{rta}
\end{equation}
where $\X = e^{\tau_{1667}}$.
The ratios for our sources range from 1.2 to 1.9, with the majority being 
around 1.8.
This confirms that we are seeing thermal emission from dark clouds and 
that the opacity is only modest. 
Equation~\eqref{rta} gives a unique solution for $\X$, which leads to 
\begin{equation}
%
%
T_x = \frac{T^{1667}_A}{1-\X}+T_B \lp
\end{equation}
Except for one source, the excitation temperature of OH ranges from
5 K to 9 K, which is not much larger than the background temperature.
At this level of excitation, 
the assumption of neglible background temperature made for 
equation~\eqref{noh} results in underestimation of $N_{OH}$ by a
factor of 1.6 to 2.3.

Because of weaker line emission at 1665 MHz and the exponential dependence 
of the antenna temperatures on the optical depth, the satellite line ratio
method is very sensitive to uncertainties in the spectra of the 
satellite component.
In some cases, the line widths of the main and the satellite line are
not the same, which may be a result of the low signal to noise ratio of
the satellite component. 
To present a more self--consistent correlation study
of HI and OH, we only use the 1667 MHz data in determining
OH column density, but note that
these results likely underestimate the OH column density by a factor 
$\simeq$ 2. 
%
%
\subsection{\c18o Column Density}

In order to determine the molecular content of dark clouds, we have chosen
as a tracer the J=1--0 transition of \c18o,
which is optically thin (or
close to it), and which has relatively uniform abundance relative to 
\h2, $\sim 1.7\times10^{-7}$ \citep{frer82}. 
The \c18o\ column density in the J=1 level can be calculated from
\begin{equation}
%
%
\frac{N^1(\text{\c18o})}{ \text{cm}^{-2}} = 3.72\times10^{14} 
[1-\frac{e^\frac{h \nu}{k T_x}-1}{e^\frac{h \nu}{k T_B}-1}]^{-1}
\frac{\int T_{mb}(V) dV}{\text{K km s$^{-1}$}}    \lc
\end{equation}
where $T_x$ is the excitation temperature of the J=1--0 transition, 
$T_B$ is the background temperature, and the integrated intensity 
is corrected for the main beam efficiency. 
The LTE approximation is then used to justify substituting $T_k$ for $T_x$ 
for all transitions in converting $N^1(\text{\c18o})$ to the total column 
density of \c18o.
The smoothing of the FCRAO (45\arcsec\ beam) data
to the Arecibo resolution gives a physical meaning to
the [H]/[\h2] thus derived: it is the ratio between the total number of
H atoms and \h2 molecules in a 3\arcmin\ beam.
%
%
%
%
\section{HI Survey Statistics and Characteristics of HINSA  \label{survey}}

\begin{figure}                                 
\epsscale{0.8}
\plotone{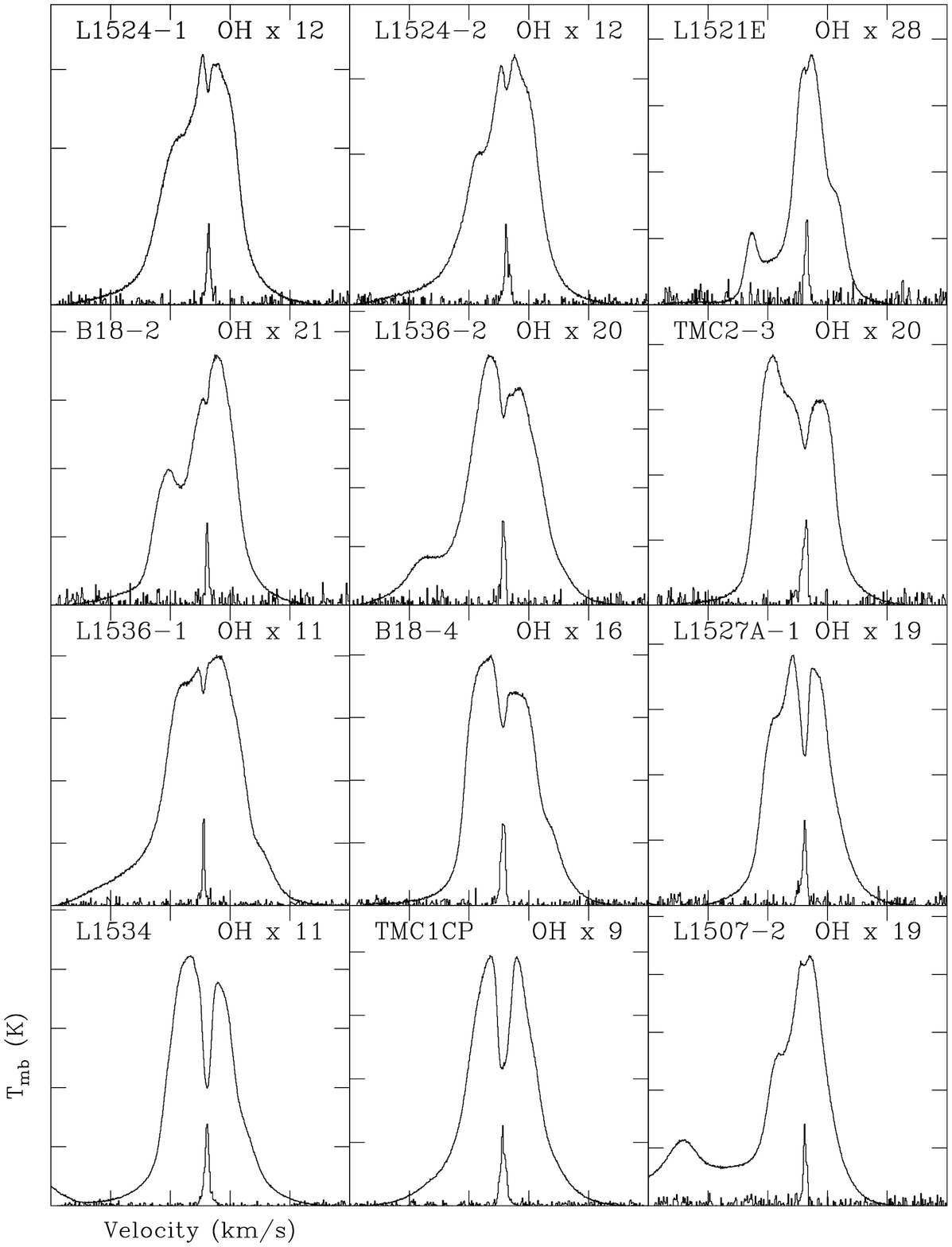}
\caption{Sources with clear HINSA features.
The axes are the same as in Figure~\ref{dipoff}.}
\label{dip1}
\end{figure}

\begin{figure}                                 
\epsscale{0.8}
\plotone{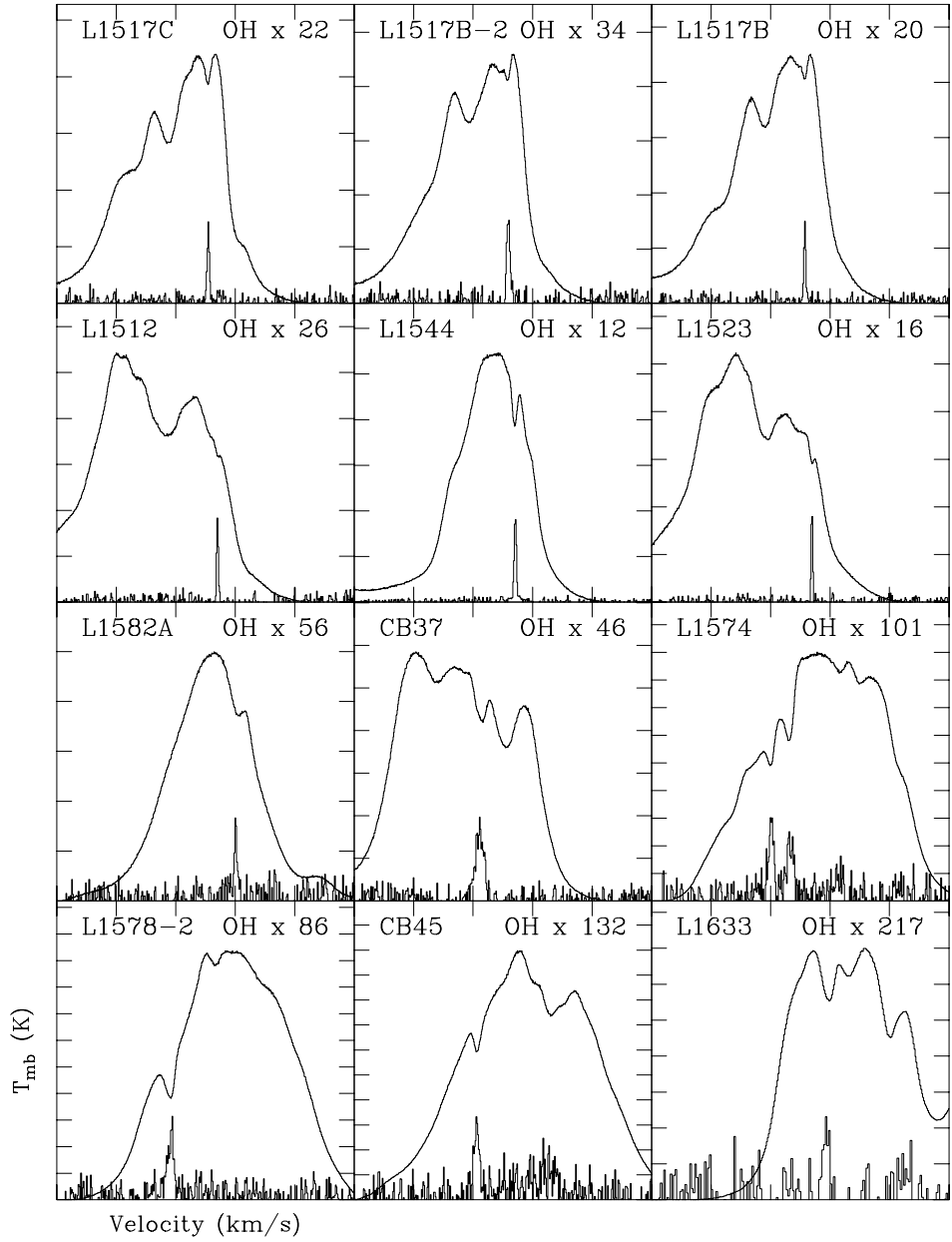}
\caption{Sources with clear HINSA (continued).}
\label{dip2}
\end{figure}

\begin{figure}                                 
\epsscale{0.8}
\plotone{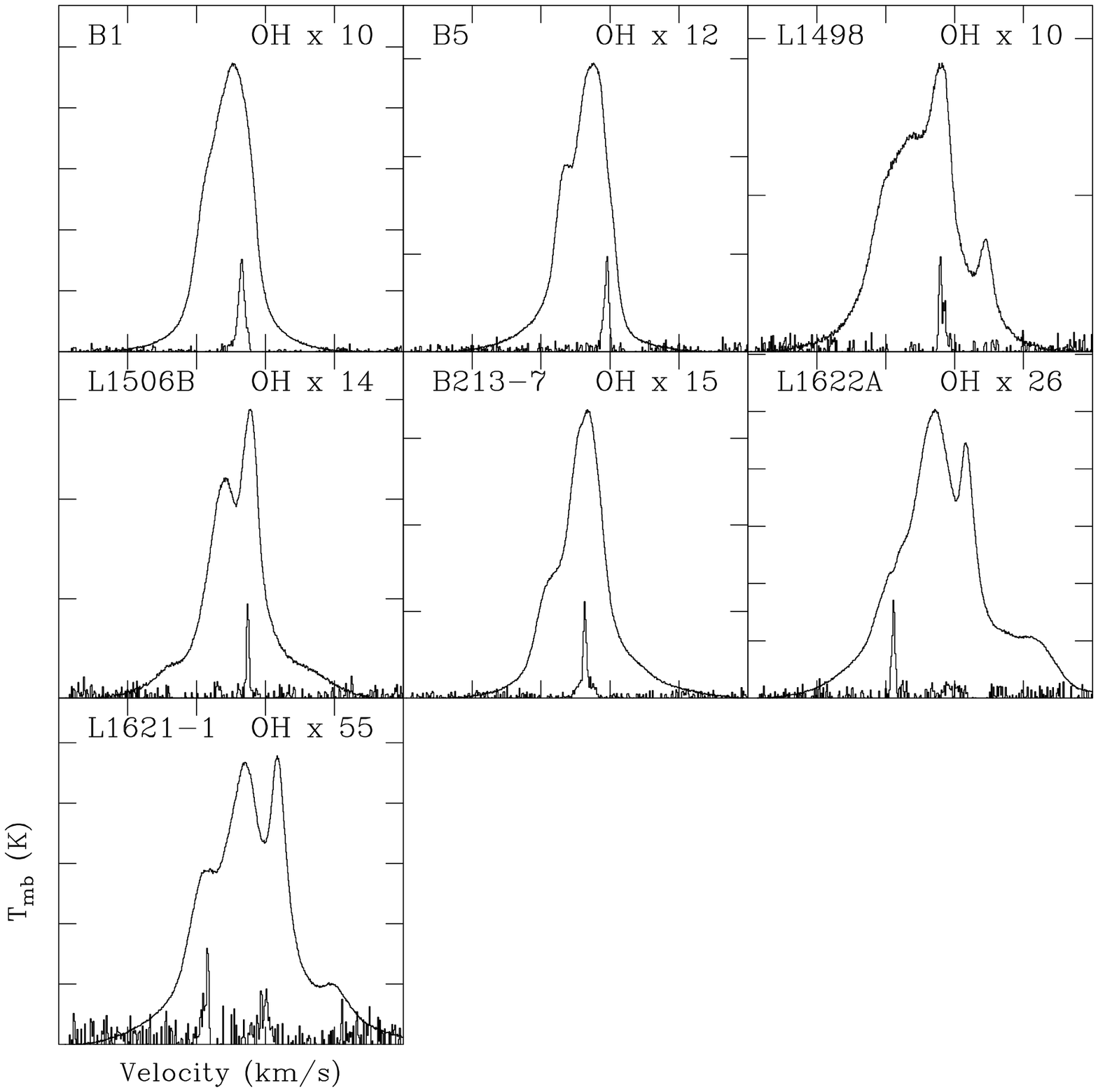}
\caption{Sources without a clear HINSA feature. 
The axes are the same as in Figure~\ref{dipoff}.
}
\label{nodip}
\end{figure}

As described in Section~\ref{obs}, we surveyed 30 nearby dark clouds,
unbiased in terms of HI absorption.
In this sample, 23 have a clear narrow absorption 
dip at the same velocity as that of the OH emission 
(Figures~\ref{dip1} \&~\ref{dip2}). 
Four more clouds (L1498, L1506B L1622A \& L1621-1) have hints of an 
absorption feature coincident in velocity with the OH emission, 
which correspond to a `shoulder' on the HI profile (Figure~\ref{nodip}). 
The remaining three clouds (B1, B5 \& B213-7) have no indication of 
narrow absorption features.
Counting the clouds with the `shoulder' features as non--detections, 
the detection rate of HINSA  toward optically 
selected dark clouds is 77 percent.
This high detection rate strongly supports the association of
HINSA with dark clouds.
Further analysis given below based on comparisons of specific tracers reinforces 
this association.

In our sample of 30 dark clouds, 7 are found to contain embedded young stellar
objets (EYSO) \citep{lee99}.  The ratio of cores with EYSO to starless cores
is thus 0.3, the same as found in these authors' larger sample.
If the detection rate of HINSA is independent of EYSO,
we would expect to find 17.6 clouds with HINSA in starless cores and 5.4 clouds
with HINSA in EYSO cores. The actual numbers in our sample are 17 and 6, 
respectively. Thus, our observations indicate that the
presence of an embedded YSO does not significantly affect
the likelihood of finding HINSA.

In our sample, the HI spectra are often structured and some 
are most likely due to multiple absorption features 
(e.g.\ L1517C, L1512, L1523, CB37, CB45 and L1633).
Out of these multiple absorption features, only those with
corresponding molecular emission are believed to be associated with
molecular clouds. 
In our sample, these features exhibit a line width 
comparable to that of \13co\ at the same velocity. 
We therefore define HI self-absorption features with corresponding CO emission 
and line width smaller that of CO as HI Narrow Self--Absorption (HINSA).
This view is compatible with the findings of cold HISA features without
molecular counterpart in the Canadian Galactic Plane Survey \citep{gibson00, knee01}. 
The Arecibo HI survey of the galactic ring \citep{kuchar94} 
also shows extended HI absorption feature without overlaying CO clouds. 
The existence of atomic hydrogen colder (15 K to 35 K) than
the standard value (80 K) for the Cold Neutral Medium (CNM), yet still
independent of molecular clouds, has also been indicated in studies of 
absorption against continuum
sources  \citep*[e.g.\ ][]{heiles01}. 
Most of the HISA is thus likely a result of temperature
fluctuations in the CNM. 
In the absence of molecular cooling, it is still 
not evident how atomic gas can be maintain at such low temperatures.
In contrast, the low temperatures of HINSA have a straightforward explanation in
terms of CO cooling if it arises in largely molecular regions. 
Clouds with HINSA are also less turbulent than typical ``neutral hydrogen'' clouds
that may produce HISA, further emphasizing the distinction between the two.
%
%
\subsection{Low  Temperature of the HINSA \label{temperature}}

There are two ways to estimate the temperature
of the atomic hydrogen detected in absorption. 

First, the line width of the absorption dip provides an upper limit 
to the kinetic temperature assuming only thermal broadening is present.
The equivalent temperature, derived from the fitted FWHM ($\Delta V$) of
HI absorption 
\begin{equation}
%
%
T_{eq}=\frac{m_H \Delta V^2 }{8\ln(2)  k}
\end{equation}
is given in Column 13 of Table 2.
There are 10 sources which have an absorption line width so narrow that
the HI responsible must be thermalized, or be very close to 
thermalization, at a temperature lower than 15 K.
The narrow line widths we have observed rule out the
possibility that the HINSA could be produced in warm (e.g. cloud halo)
gas with T $\simeq$ 100 K, even allowing for modestly subthermal 
excitation of the 21cm transition.

Subtle complications in obtaining line parameters from
an absorption feature arise when the dip is on the slope of the
background emission. 
In general, the fitted absorption line appears to peak closer to the 
center of the emission feature, and to appear narrower than it really is. 
The corrections required in both cases are small compared to the line 
width of the narrow line HI absorption.
We do not correct for the displacement of velocity peak, since the 
average effect is zero due to the randomness of the dark cloud velocity 
with respect to that of the background HI emission.
The correction for line width is on the order of 0.04 \kms\ \citep{lev80}, 
which is only significant when calculating $T_{eq}$ for a couple of unusually 
narrow line width sources. 
The uncertainty thus introduced in $T_{eq}$ for L1523, L1521E and L1517B 
is between 0.5 K $\sim$ 1 K. 
For other sources, this correction can be neglected.

Second, the excitation temperature (or the spin temperature for the 21cm line)
can be estimated by rewriting equation~\eqref{tab}
\begin{equation}
%
%
	T_x = T_c + [pT_{HI} - \frac{T_{ab}}{1-e^{-\tau}}]/(1-\tau_f) \lp	
\end{equation}
The optical depth of the narrow line absorbing gas is on the order of
unity. 
Substituting an infinite $\tau$ into the equation above overestimates
$T_x$. 
For the optical depth of the foreground emission $\tau_f$,
a nominal value of 0.1 is used, which is probably an overestimate 
obtained by assuming the total optical depth
of galactic HI emission ($\tau_h$) to be unity.
The upper limit of HINSA excitation temperature is thus
\begin{equation}
%
%
T^{upper}_x = T_c + [pT_{HI} - T_{ab}]/0.9 \lp
\end{equation}
These upper limits are listed in Column 14 of Table 2.
Compared to the temperature limit set by the line width, this method
has relatively large uncertainties resulting from the assumptions about
$p$ and the optical depth.
Nonetheless, it provides an independent estimate based on the depth of the
HINSA profile.
It also confirms our understanding that the majority of HINSA features
come from cold gas ($\lesssim 40$ K) with some sources in the range 
of 10 K.

These two estimates of temperature upper limits reveal
HINSA to be associated with cold gas. 
According to the lower value between the two upper limits, the average
temperature upper limit to the 24 HINSA features is 23 K.
Moreover, a significant fraction of the sources must have atomic gas as cold as 7--15 K.
The well shielded regions of molecular cloud cores are the most likely, 
if not the only, sites that can plausibly contain material this cold.
%
%

\subsection{HI and OH Correlation}

\begin{figure}[htp]                                 
 \subfigure{\includegraphics[width=.60\textwidth, angle=90]{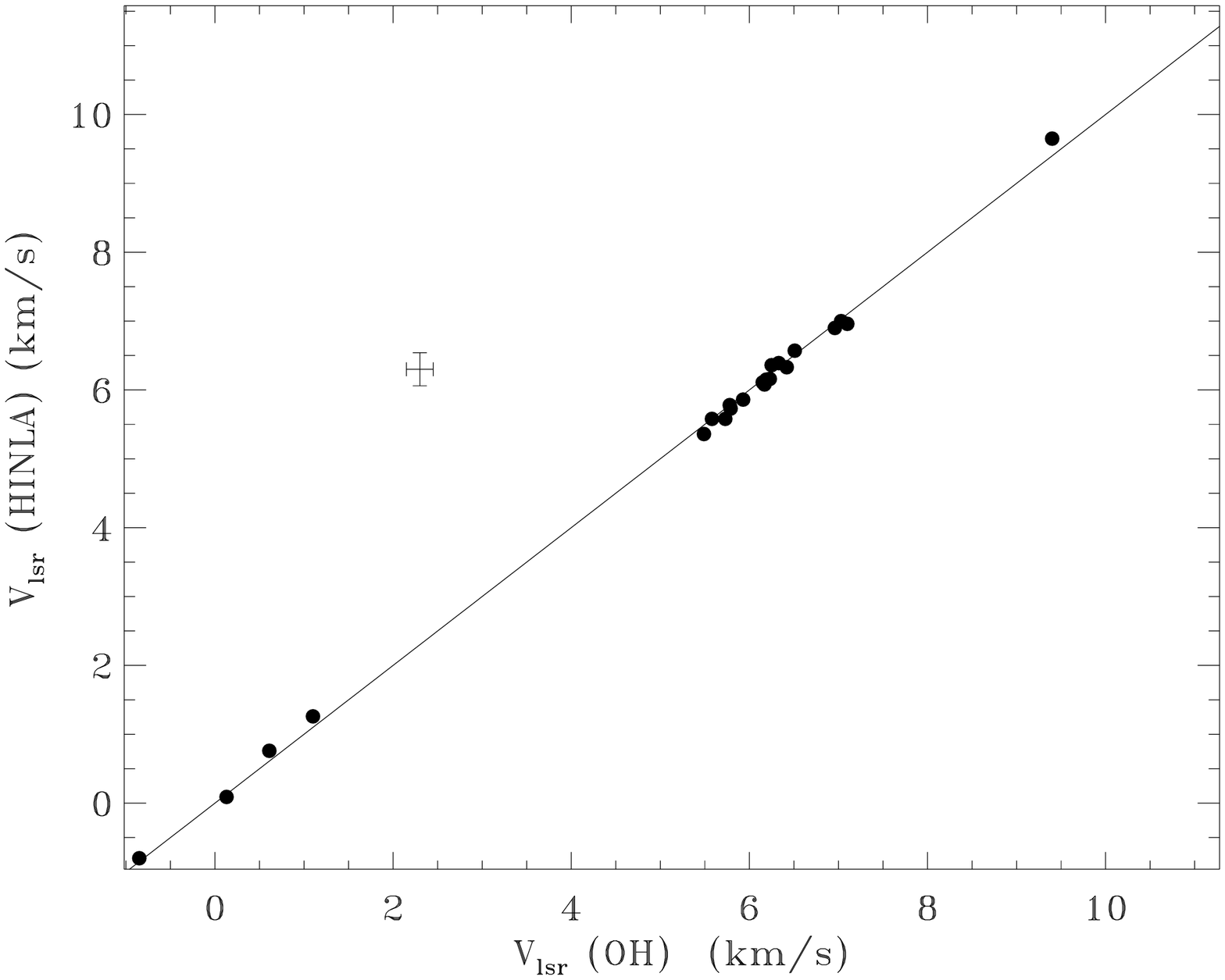}}\\
     \subfigure{\includegraphics[width=.60\textwidth, angle=90]{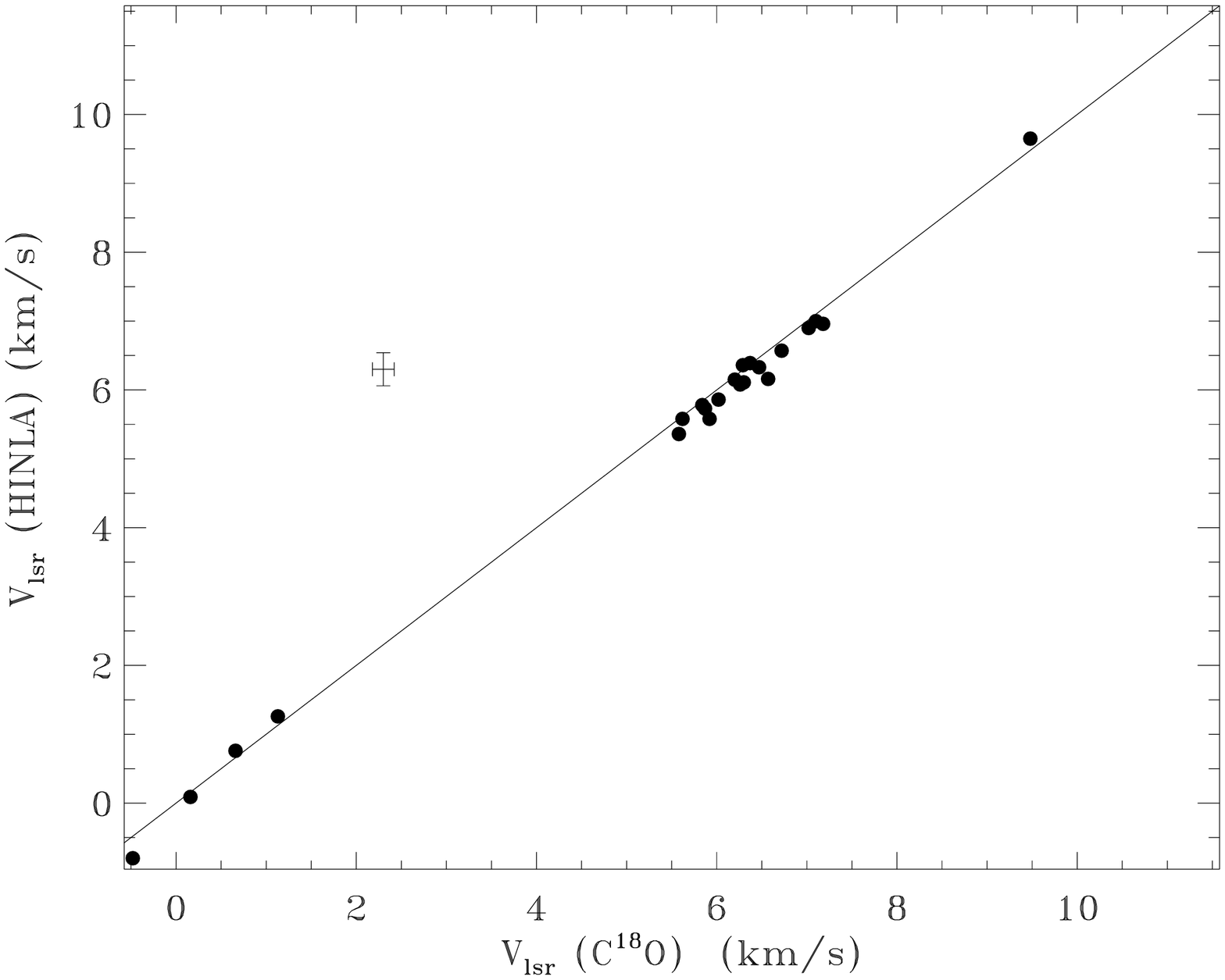}}
\caption{Upper panel: centroid velocity
 (relative to the local standard of rest) of HINSA vs. that of
the OH 1667 MHz emission. Lower panel: centroid velocity of HINSA vs.
that of the \c18o. 
The error bar represents 3 $\sigma$ statistical uncertainties. The root mean
square difference in the velocities is 0.03 and 0.02 \kms\ for
the upper and lower plots, respectively.}
\label{fig:vlsr}
\end{figure}

For sources with well defined absorption features, the
similarities between the OH and HINSA profiles are obvious. 
The two profiles are coincident in velocity as shown in Figure~\ref{fig:vlsr}.

The line widths of different species are important indicators of
the conditions in the regions they are found, as well as of the extent
of the regions responsible for the spectral line in question.
The well known empirical velocity--line width relations all suggest larger 
line widths at larger spatial scales (e.g.~Larson 1981;  
Caselli \& Myers 1995).
To compare the line widths of species with very different molecular/atomic 
weight, it is appropriate to separate the thermal from the nonthermal
contributions to the line widths.
This can be done by defining the nonthermal line width as
\begin{equation}
\label{vnt}
%
%
\Delta V_{nt} = \sqrt{\Delta V^2 - \frac{8 \ln(2) k T_k}{\mu  m_H} } \lc
\end{equation}
where $\Delta V$ is the observed FWHM of the line, $\mu$ is the molecular 
or atomic weight, $m_H$ is the hydrogen mass and $k$ is Boltzmann's constant. 
For  HINSA,  the width is that of the Gaussian fitted to the absorption
profile, while for the other species we fit the emission spectra.
The correction for the thermal broadening is significant  
for the HI due to its low mass.
The nonthermal line width gives a better indicator of the 
spatial extent of its progenitor, since the mass dependence of the 
thermal broadening has been corrected for. 
The line width data can be found in Table 2.

It is evident that the HINSA, OH, and \c18o line widths are well
correlated, while the CI exhibits quite different behavior.
The average nonthermal line width of the HINSA  and of the OH 
are 0.85 \kms and 0.83 \kms, respectively. 
They are essentially the same as the average line width of \13co at 0.82 \kms. 
These results indicate that the HINSA is produced in regions of
significant extinction where the molecular abundances are appreciable.

We have obtained direct information on the spatial correlation of HINSA and
OH emission by extensive mapping of L1544, discussed
below in Section \ref{l1544}.
Limited evidence for spatial correlation of OH and HINSA can also be found
by looking at the OFF source spectra (Figure~\ref{dipoff}).
When the telescope beam is moved away from the center
of each source, it is evident that 
the HINSA absorption dip starts to disappear along with the OH emission. 
There is no hint of any increase in the strength of the absorption 
features at the cloud boundaries, as would be expected if the gas
producing the HINSA were in some peripheral zone of the clouds we 
have studied \footnote{We hesitate to use the term `limb brightening' 
to describe an increase in the depth of the absorption line, but
if the HI in question were in an outer `onion skin' of the cloud,
a map of the integrated area of the absorption would presumably 
exhibit this behavior.}.
%
%
\subsection{Correlation with CO Isotopologues and CI}

As listed in Table 2, all observed species, HINSA, OH, \c18o, 
and CI, have essentially the same velocity relative to the local
standard of rest ($V_{lsr}$) in a given cloud. 
The V$_{lsr}$ of the HINSA and of the \c18o\ are plotted in 
Figure~\ref{fig:vlsr}. 
The root mean square separation of our data points from the line of 
equal velocities is 0.03 \kms, which is small compared to both the 
line width and the velocity resolution.

As traced by the line widths, the spatial distributions within the clouds
of the CO isopologues, the HINSA, and the HI appear to be somewhat different. 
As mentioned above, HI, OH, and \13co\ have, on average, the same line 
width.
\c18o has the narrowest line width ($<\Delta V>$ $\sim$ 0.64 \kms), 
presumably tracing the innermost core. 
The CO line is wider with $<\Delta V>$ = $1.2$ \kms, while
the CI line is much wider still with  $<\Delta V>$ = 2.3 \kms. 
The large nonthermal line width of the CI confirms this species to be 
a tracer of PDR regions, where more turbulence exists. 
HINSA seems to be mixed with relatively quiescent molecular material
at higher extinction.

\begin{figure}[htp]
     \centering 
     \subfigure{\includegraphics[width=6cm, angle=90]{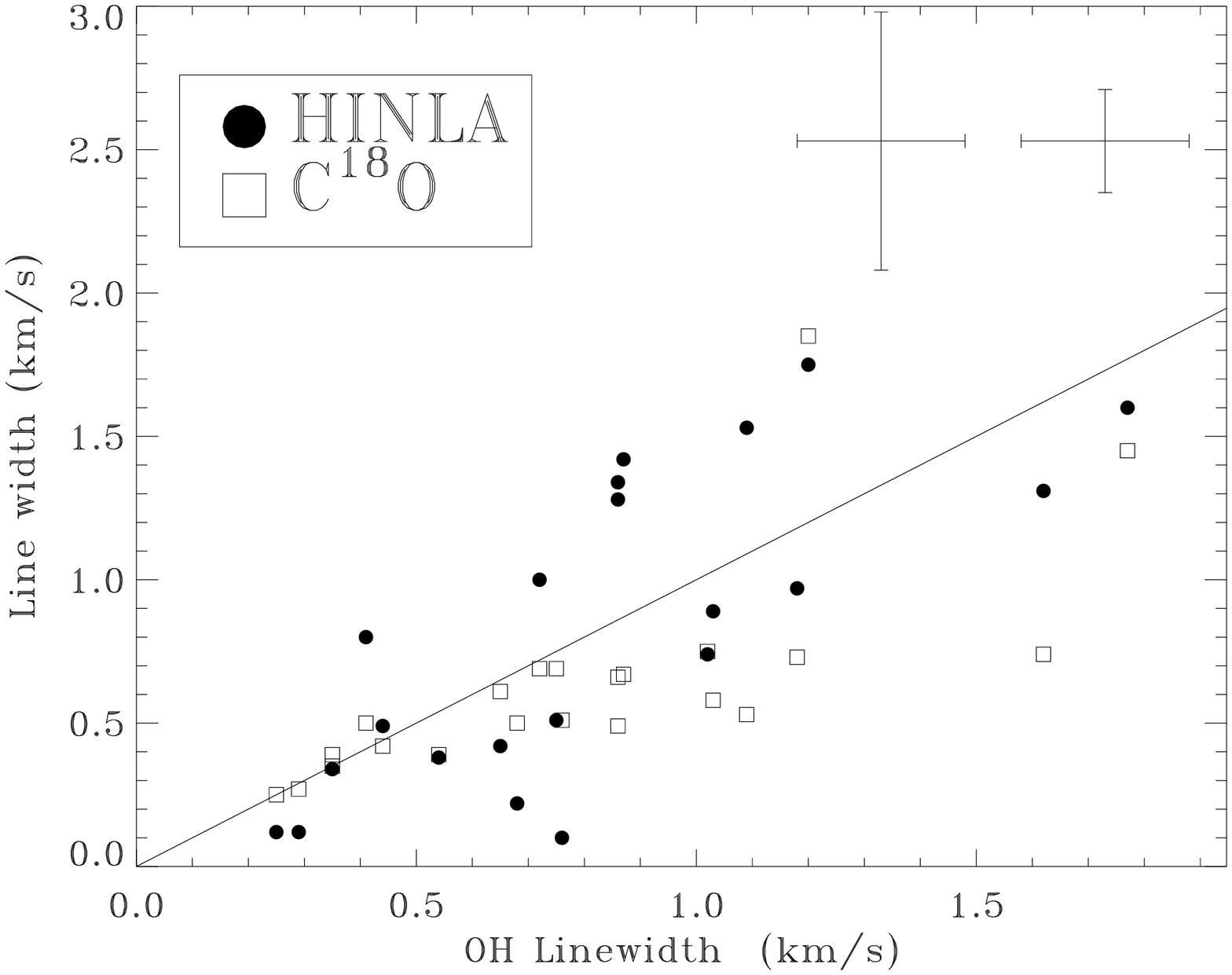}}
     \subfigure{\includegraphics[width=6cm, angle=90]{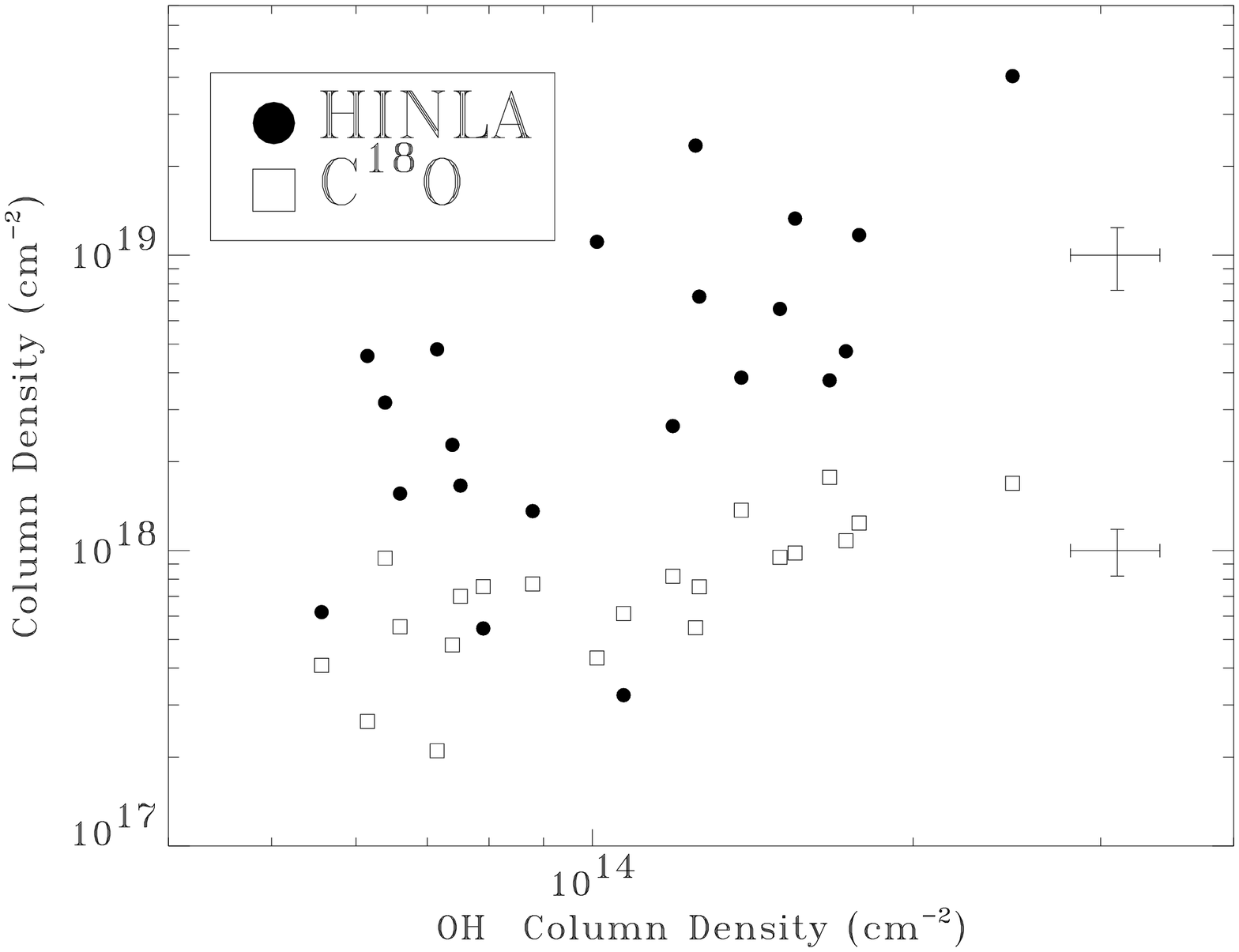}}\\
     \subfigure{\includegraphics[width=6cm,angle=90]{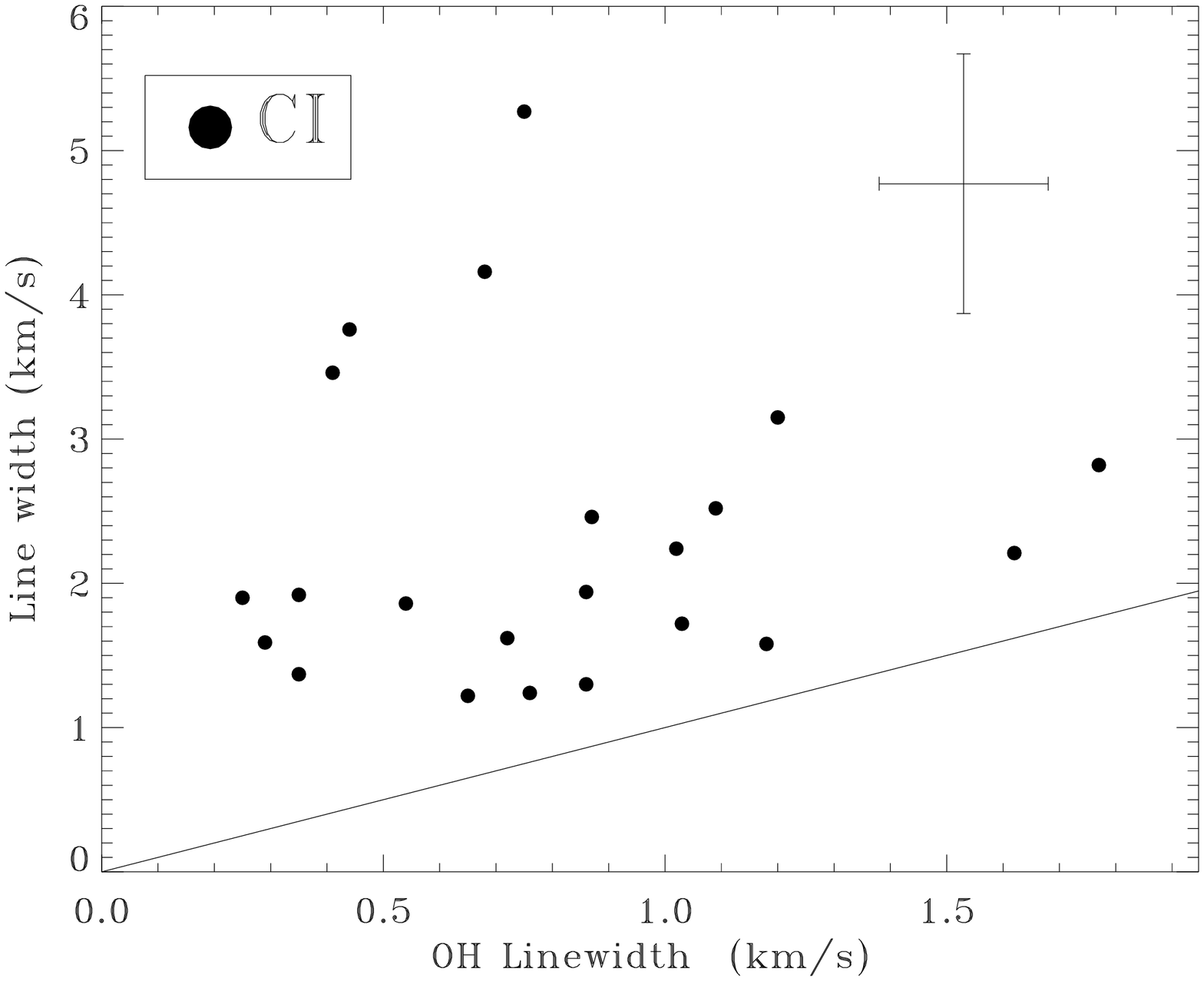}}
     \subfigure{\includegraphics[width=6cm, angle=90]{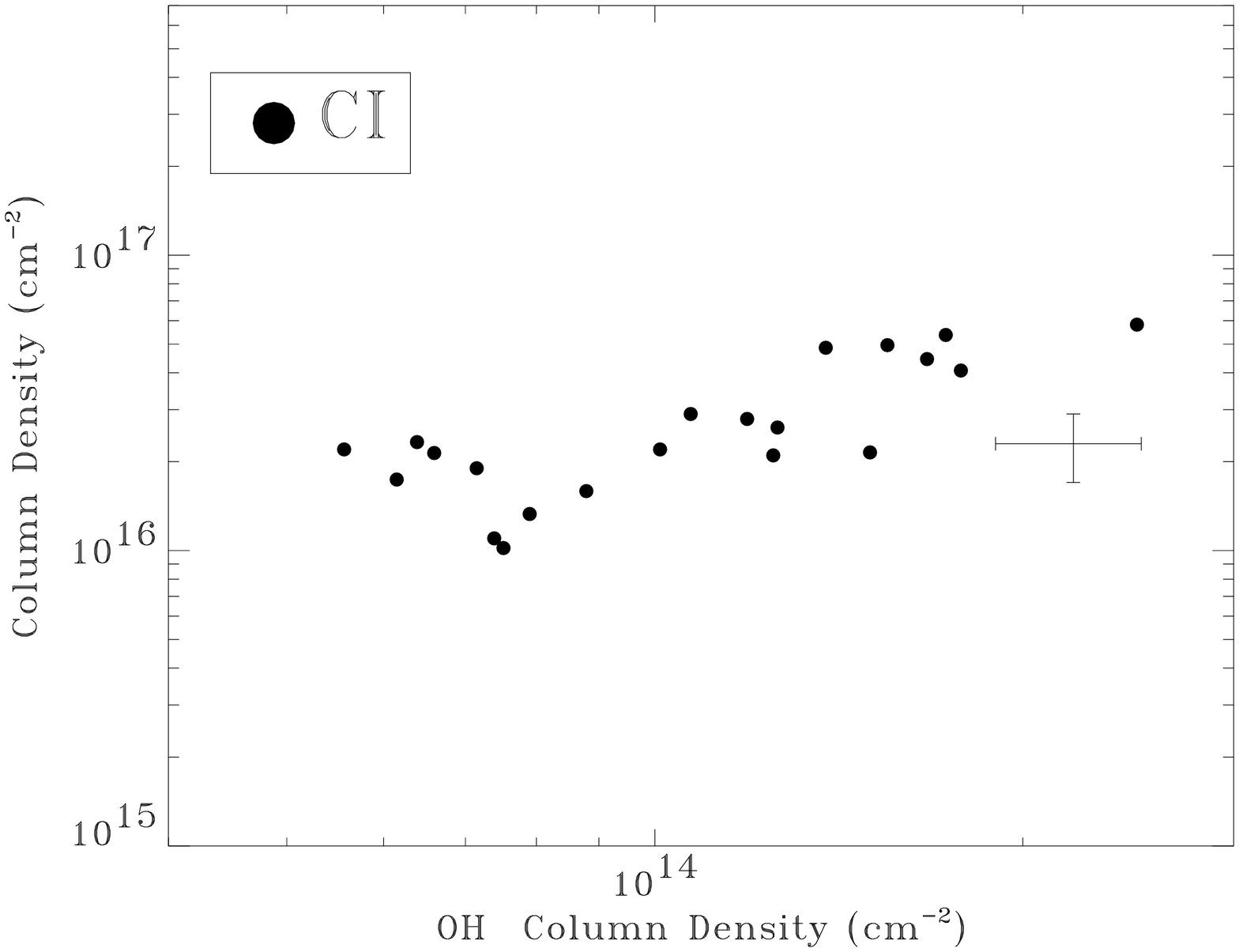}}
\caption{Left panels: nonthermal line width of OH, \c18o, HINSA and CI.
Right panels: column densities of of OH, \c18o, HINSA and CI. 
In the upper right panel, the values of \c18o\ column density have been 
multiplied by a factor of 10$^3$ in order to be plotted on the same scale.
The straight lines in the left hand panels indicate equal line widths
for the species observed and the OH, used as a reference.  The HINSA
and \c18o emission exhibit essentially the same line width as the OH
in a given cloud, while the CI line width is significantly larger than 
that of the OH. The error bars in all the panels represent 3$\sigma$ 
errors. The uncertainty level is given in Table 2 and 3.}
\label{fig:col}
\end{figure}

For individual sources, a positive correlation exists between the line width of 
OH and of HINSA, as shown in the upper left panel of Figure~\ref{fig:col}. 
The probability of the null hypothesis (the two being uncorrelated)
is $r$ = 3.6$\times10^{-5}$; the linear correlation coefficient is 0.76. 
A similar correlation exists between the OH line width and that of \c18o. 
In contrast, no such correlation exist between the OH and CI line widths. 
Again assuming a linear correlation, the null hypothesis is
very likely ($r=0.49$).

For the column densities, only OH and \c18o\ are definitely
positively correlated. 
For HINSA, the correlation is not clear. This is somewhat
expected from the standard \h2\ formation model (section~\ref{nhih2}),
as the column density of HINSA should be correlated rather with the cloud
size than with that of \h2. 
For CI, the assumption of small optical depth has not been tested in 
our study and thus may cause large uncertainties. The current 
data set does not definitively answer all questions about the correlation
between column densities of the different species, a subject which
deserves additional consideration.
%
%
\subsection{L1544}
\label{l1544}

To get a better idea of the spatial correlation of HINSA and 
molecular tracers, we mapped the quiescent dark cloud L1544. 
This object has relatively narrow line widths in molecular tracers 
($\Delta V(\text{CO}) = 1.0$ \kms; $\Delta V(\text{C}^{18}\text{O}) = 0.44$ \kms).
It is usually characterized as starless, due to there being no clear
association with an IR source \citep{ward94}. 
The inner 5\arcmin\ core region around the center of L1544
has been well studied and found to exhibit molecular depletion and infall
motions \citep{case02, tafa98}. 

\begin{figure}[htp]
     \centering 
      \includegraphics[width=.70\textwidth, angle=-90]{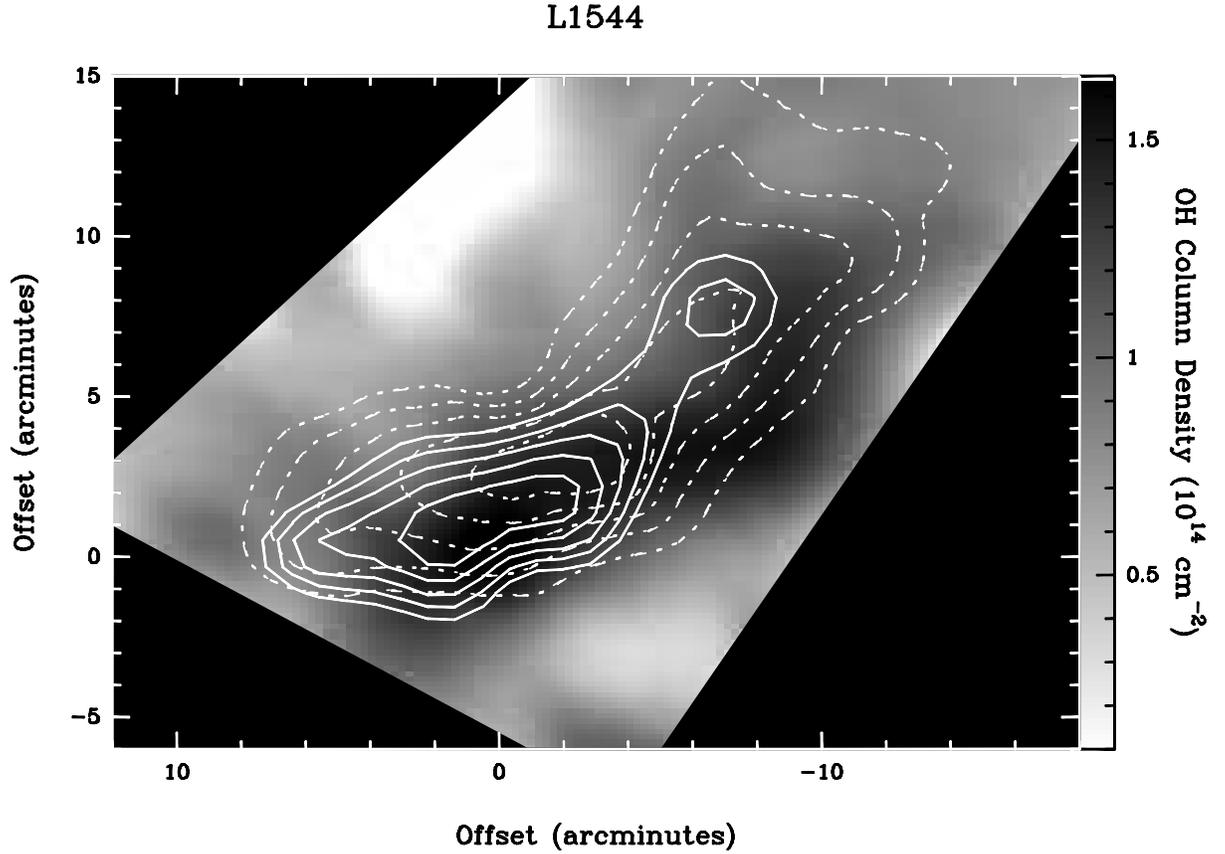}
\caption{Column densities of OH (grey scale) overlaid with column densities 
of HINSA (dashed contours) and integrated intensities of \c18o\ 
(solid contours). 
Both HINSA and \c18o\ contours are at 50$\%$,  60$\%$,  70$\%$, 80$\%$, 
and 90$\%$ of their respective peak values. 
The \c18o data are from \citet{tafa98} and have been smoothed to the 
3\arcmin\ Arecibo resolution.}  
\label{fig:l1544}
\end{figure}

We mapped an extended region of L1544, which is the same
area as outlined by extinction \citep{snell81}. 
The RMS of our HI and OH spectra are smaller than 0.1 K. 
At the cloud periphery, the OH 1667 MHz emission is generally weaker 
than 3 $\sigma$.
All three tracers, OH, HINSA, and \c18o\ delineate the same cloud 
(Figure~\ref{fig:l1544}). 
The \h2 column density in the center of this cloud derived from \c18o 
and assumed standard fractional abundance is 6.5$\times10^{21}$ \cm2.
Taking dimensions of the cloud core of 9\arcmin\ by 16\arcmin\ gives
a characteristic size of 0.49 pc at an assumed distance of 140 pc.
This results in a characteristic volume density 
$n$(\h2) = 4.3$\times10^3$ \cc.

For a spherical cloud of uniform density in virial equilibrium ignoring
magnetic fields and external pressure, the \h2 density is given by
\begin{equation}
	\frac{n_v(\text{H}_2)}{\text{cm}^{-3}} = 2.03\times10^3 
			\frac{1}{\mu} 
			[\frac{\Delta V /(\text{km s}^{-1})}{R/(\text{pc})}]^2 \lc
\end{equation}
where $\Delta V$ is the FWHM line width, $R$ is the cloud radius, and $\mu$ is
the mean molecular weight.
Substituting the appropriate values yields $n_v(\text{H}_2) = 2.9\times10^3$ \cc.
This is in reasonable agreement with the average density derived above,
and suggests that this cloud is in, or not far, from virial equilibrium.

The secondary maximum in OH located around offsets (-8\arcmin, 3\arcmin)
is due to a localized increase in the line width, indicating an increase
in turbulence. 
At the center of the cloud, the line width of the 1667 MHz
OH emission is 0.47 \kms, while at (-8\arcmin, 3\arcmin),
it is 0.70 \kms. 
This second, more turbulent peak in the OH column density
does not correspond to an enhancement in either \c18o\ or HINSA.
A likely scenario is that the OH traces a low extinction envelope,
which exhibits fluctuations in MHD or some other kind of turbulence,
while \c18o\ represents
 the total column density of quiescent
material. 
HINSA produced by cosmic ray dissociation (see Section \ref{nhih2} 
below), which is unaffected by extinction, should also be able to
trace the highest column density regions. 
On the other hand, if HINSA is produced mainly by photodissociation
in the envelope, a limb brightening effect should appear. 
Such an effect is not seen for L1544. 

A recent study has also indicated a correlation between HI self--absorption
and molecular emission for the molecular cloud GRSMC 45.6+0.3 
\citep{jackson02}. That study, however, suggests that HI absorption
arises from cloud skins with $A_v<2$.
Even at its near distance of 1.8 kpc, this source is more than
an oder of magnitude more distant that of L1544, so the detailed location 
of the cold HI relative to the molecular material is hard to be determined
with confidence. 
Our study of L1544 favors a ``distributed'' rather than an ``external skin''
layer for HI, but this conclusion should be confirmed by additional detailed
maps of nearby clouds.
%
%
%
%
\section{The Atomic Hydrogen Density in Molecular Clouds \label{nhih2}}
 	
The majority of molecular hydrogen in the ISM is thought to be formed 
by reactions on dust grains.  
This process is also the major destruction pathway for hydrogen atoms. 
The steady state \h2\ production rate (\cc\ \s) $R_{H_2}$ can be  written as 
\begin{equation}
%
%
R_{H_2}=0.5 n_g n_1 \sigma v S \eta \lc
\label{rh2}
\end{equation}
where $n_g$ is the grain number density, $n_1$ is the HI density in the 
gas phase, $\sigma$ is the grain cross section, $v$ is the relative velocity 
between H atom and grains (essentially the velocity of H), $S$ is the 
sticking probability of an H atom on a grain, and $\eta$ is the 
formation efficiency. 
The total density of hydrogen atoms is $n_H$ = $(n_1+2n_2)$, where 
$n_2$ is the density of hydrogen molecules.
The grain and proton number densities are related through $n_g = gn_H$. 
The value of $g$ depends on the grain model and gas conditions.
A `standard' dust grain is taken to have a radius of 0.1 \mic\ and a 
density of 3 g \cc.
In a molecular cloud, we can reasonably assume the following gas conditions:
[He]/[H] equal to 0.09, most hydrogen gas in the form of \h2, 
and an gas to dust mass ratio of 100. 
Based on these assumptions about the dust grains and the gas,
we determine $g$ to be $1.8\times10^{-12}$.

In the standard picture of \h2 formation (Hollenbach \& Salpeter 1970), 
the time scale for H atoms deposited on the grain to `cover' the 
grain surface through tunneling processes is much shorter than the 
residence time of an adsorbed H atom. 
Since the \h2\ formation reaction is exothermic (4.5 ev released), 
$\eta$ is taken to be 1 in their model as long as there are more than two 
H atoms on a grain. 
Equation \ref{rh2} with a near unity $\eta$ works well in 
diffuse clouds with S$\sim$0.3 \citep{holl71, jura75}. 
A recent study of sticking probability on icy surfaces gives $S$ close 
to unity at low temperatures \citep{buch91}. A recent analysis including
both chemisorption and physisorption on grain surfaces indicates that
\h2\ formation can be efficient over a wide range of temperatures
\citep{cazaux02}, specifically including the 8 K - 20 K range of interest
for dark clouds.
Adopting $S = 0.5$ and $\eta$ = 1.0 for the following discussion, we obtain  
the equation 
\begin{equation}
%
%
R_{H_2}=2.06\times10^{-18}n_H n_1 \sqrt{T} \lc
\end{equation}
where $T$ is the gas temperature, for the formation of rate of \h2.

For typical dark cloud conditions with temperature of 10 K, 
$n_H$ $\sim$ $10^{4}$ \cc, the HI to \h2 conversion time scale, 
$n_1$/R$_{H_2}$, is approximately 0.5 million years. 
Such a rapid conversion leaves an essentially `molecular' cloud in 
which the atomic component is maintained by cosmic rays in a steady state.
The destruction rate of \h2 is $\xi$\n2\ (\cc\ \s), where $\xi = 3\times
10^{-17}$ \s\ is the cosmic ray ionization rate. 
This parameter has approximately a factor of 3 uncertainty associated
with it, as indicated by the varied results and range of fits obtained
for different sources and models by e.g. \cite{case98b}; \cite{case02}.
In a steady state and assuming \n2$\gg$ \no,
\begin{equation}
%
%
\dot{n_2}=2.06\times10^{-18}\times 2n_2 n_1 T^{1/2} - \xi n_2=0 \lc
\end{equation}
which, for 10 K, gives \no\ to be  2.3 \cc.

\begin{figure}[htp]
\includegraphics[width=.80\textwidth, angle=90]{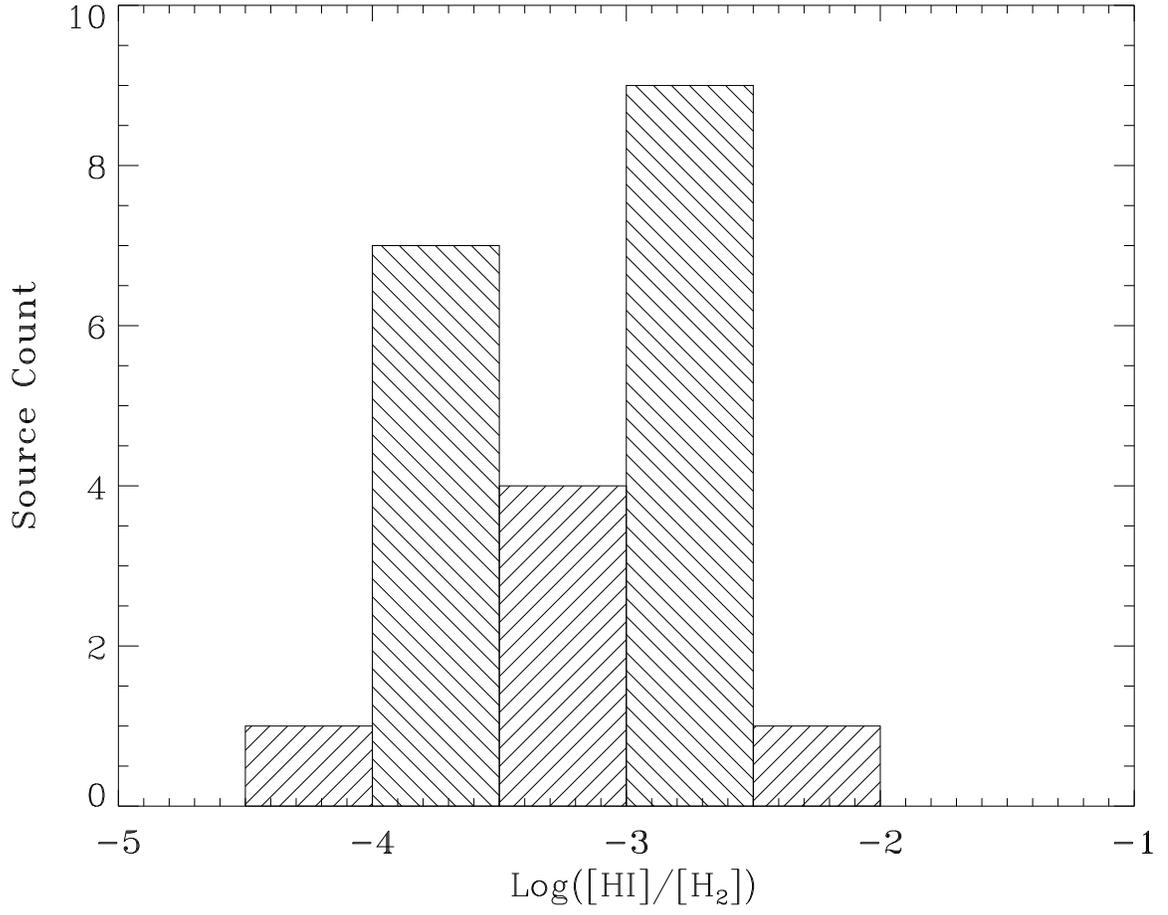}
\caption{The distribution of the abundance ratio [HI]/[H$_2$] in 22 dark clouds.}  
\label{fig:abundance}
\end{figure}

This HI density is {\em independent} of the local gas density 
\footnote{
This is by no means a new result; e.g. \cite{solo71} showed
that a constant $n_1$ was to be expected, although their value was more than
an order of magnitude larger due to the large value of the cosmic
ray ionization rate which they adopted.}.
Therefore the entire region containing molecular material can
contain cold HI and be capable of producing HINSA, with essentially
an equal contribution per cm of path along the line of sight.
In our sample of clouds, the average column density of HINSA is 
7.2$\times10^{18}$ \cm2. 
We adopt an angular size for the region with absorbing HI of 15\arcmin, 
which is consistent with the typical size of Lynds clouds and Bok globules
in the Taurus complex.
This corresponds to a physical dimension equal to 0.6 pc at a distance
of 140 pc.
The average \h2\ gas density which we derive from the average 
column density of \c18o and fractional abundance 1.7$\times10^{-7}$
is equal to 2.5$\times10^3$\ \cc.  
The average HI fractional abundance [HI]/[\h2] in the dense molecular gas
predicted from the standard theory is 9.2$\times10^{-4}$.
This is in reasonable agreement with the observed result 1.5$\times10^{-3}$
(see Table 3 and Figure~\ref{fig:abundance}).

Another obvious process for producing atomic hydrogen associated
with molecular clouds is photodissociation. 
In a plane parallel PDR model with microturbulence \citep{wolfire93, 
jackson02}, a large fractional abundance of HI ($>10^{-2}$) can be 
produced at low extinctions ($A_v<2$). 
We do not see this HI envelope, either in total column density
or in morphology. 
One possible explanation is that the conditions in such regions are much
less favorable for production of HINSA than the cold, quiescent
regions of the clouds.
If we ignore foreground and background absorption as well as any
continuum, we see from equations~\eqref{tab} 
that the magnitude of the absorption line from the optically thin, thermalized
gas responsible for the HINSA in the cloud is
\begin{equation}
\label{HINSAtau}
%
%
\frac{T^0_{ab}}{\text{K}} = 5.3\times10^{-19}\frac{N(\text{HI})/{(\text{cm}^{-2})}}{\Delta V/(\text{km s}^{-1})}[\frac{T_{HI} - T_x}{T_x}] \lc
\end{equation}
where $T_x$ is the excitation temperature of the atomic hydrogen.
As the temperature in the PDR region rises, so does $T_x$, and the 
absorption intensity drops; as $T_x$ approaches the background 
temperature, the drop becomes precipitous.
In these externally ionized and heated regions, it is reasonable that the
turbulence, and hence the line width, is considerably greater than
in the quiescent cloud material, which also weakens the absorption per
hydrogen atom.

As pointed out by \citet{wolfire93}, the microturbulent
model has difficulty in reproducing CO 1--0 line profiles.
Another scenario is clumpy cloud models with macroturbulence.
The difficulty with such models is that the observed line
profiles are usually very smooth, which is hard
to reproduce by a clumpy structure.
It would be very interesting to see the predictions
of PDR models and radiative transfer calculations with numerous 
tiny clumps.
%
%
\section{Discussion \label{discussion}}

Through observations of HINSA, we have identified cold atomic hydrogen 
associated with molecular clouds. 
A detailed explanation of HINSA and related molecular species would require
comprehensive models, which may involve PDR, clumpy structure, gas and 
grain surface chemistry \citep{ruffle00}, and possibly other effects such
as turbulent diffusion \citep{willacy02}. 
An accurate determination of the abundance of cold HI and its spatial 
distribution thus constitutes a severe test of the chemistry 
and physics of `molecular' clouds.

A steady state calculation using standard cosmic ray
rate and \h2\ formation rate is shown to produce approximately the
amount of cold HI observed. 
The data do suggest, however, that there may be a need for modest
additional sources of atomic hydrogen in cold molecular regions.

The simple steady state model may not be an accurate or complete
picture, however; with improved confidence in the \h2 formation rate,
the HI fractional abundance we have
observed can be used to put an
upper age limit on these clouds. 
The question of whether the \h2\ formation rate does slow down in dense 
cores can be answered by observations at higher spatial resolution 
($<1$\arcmin), which will either identify HINSA features correlated with
high density tracers, or show HINSA to be smooth on such scales.

HINSA traces different population of neutral hydrogen from that
constituting the warm HI halos also associated with molecular clouds 
\citep{andersson91}.
The HI halos are mostly seen by virtue of the enhancement of HI emission
around molecular clouds. 
Statistically, the HI emission maxima have been shown to lie outside 
the clouds as defined by their CO emission \citep{wannier91}.
Such halos have to be warm to be seen in HI emission with
the galactic background. In fact, the absorption measurements
against continuum sources place the halo temperature around B5
to be 70 K \citep{andersson92}, well above the temperatures of HINSA.
HI halos are also shown to be much more spatially extended
than is the CO emission \citep{andersson91}. 
This is also in contrast with our observations of HINSA, which place cold 
HI inside CO clouds, a region region of size between those characteristic
of \c18o\ and \13co\ emission.

%
%
\section{Conclusions}

We have surveyed 31 dark clouds using Arecibo, FCRAO, and
SWAS. The analysis of these data show  the following.
\begin{enumerate}

\item The  21cm HI narrow self--absorption (having line width smaller 
than that of the corresponding CO emission) is
 a widespread phenomenon, detected in $\simeq$ 77$\%$ of 
our sample of dark clouds in the Taurus/Perseus region.
We use the term HINSA to distinguish the narrow absorption definitely caused
by molecular cooling from
broader absorption features seen in other surveys of HI throughout
the Galaxy.

\item The atomic hydrogen producing the HINSA absorption has
significant column density with $N(\text{HINSA}) \sim 7\times 10^{18}$ \cm2.

\item The gas responsible for the HINSA is at low temperatures, between 
10 and 25 K. Some sources (L1521E, L1512, L1523) 
must be thermalized at temperatures close to or lower than 10 K.

\item The nonthermal line width of HINSA is comparable to 
the line width of \13co, only slightly larger than that of
\c18o, and is smaller than the line widths of CO or CI.
This suggests that HINSA is produced by cold atomic hydrogen in regions
of moderate extinctions with $A_v$ larger than a few.

\item In the maps of L1544, HINSA is morphologically similar to \c18o.

\item The low temperature, the absence of increased absorption at cloud
edges, and the narrow line width of HINSA suggest that the atomic
hydrogen producing HINSA is mixed with the gas in cold, well--shielded
regions of molecular clouds.
\end{enumerate} 

%
%
\acknowledgments
The National Astronomy and
Ionosphere Center is operated by Cornell University under a Cooperative
Agreement with the National Science Foundation.  The Five College Radio
Astronomy Observatory is supported by NSF grant AST97-25951. 
SWAS operations are supported by NASA contract NAS5-30702.
We thank M. Tafalla for letting us  use his \c18o data on L1544.
Numerous discussions with P. Myers and F. Bensch provided us with
valuable insights. We gratefully acknowledge T. Bania for pointing out
important references. We thank X. Zhuo for her kind help with illustrations,
and the anonymous referee for a number of suggestions which helped improve
the paper.
%
%
%
\newpage

%
%
%

%
%

\clearpage
\clearpage

\clearpage

\clearpage

\clearpage

\clearpage

\clearpage

\clearpage

\clearpage
\small
\begin{deluxetable}{lcccc}
\tablewidth{0pt}
\tablecaption{HI Studies of Dark Clouds--Source List}
\tablehead{
\colhead{Source Name}  &\colhead{RA}          & \colhead{DEC}  &\colhead{HINSA Detection} & \colhead{EYSO Presence} }
\startdata
B1	&  033012.0	& 305726	&N	&Y  \\ 
B5	&  034428.7	& 324433	&N	&N  \\ 
L1498	&  040751.8	& 250135	&?	&N  \\	
L1506B	&  041603.9	& 251300	&?	&N  \\
B213-7	&  042212.8	& 262610	&N	&N  \\
L1524-1&  042617.8	& 243126	&Y	&Y  \\
L1524-2&  042616.9	& 242502	&Y	&Y  \\ 
L1521E	&  042617.2	& 260734	&Y	& N \\
B18-2	&  042934.4	& 244546	&Y	& N \\
L1536-2&  042950.7	& 225329	&Y	&N  \\
TMC2-3	&  042957.6	& 241126	&Y	& N \\
L1536-1&  043019.7	& 223651	&Y	&N  \\ 
B18-4	&  043234.0	& 240248	&Y	& Y \\	
L1527A-1& 043505.1	& 260837	&Y	& Y \\ 
L1534	&  043636.4	& 253514	&Y	& Y \\
TMC1CP  &  043838.5	& 253630	&Y	&N \\
L1507-2&  043953.4	& 293825	&Y	&N  \\
L1517C	&  045133.6	& 303054	&Y	& N \\
L1517B-2& 045156.4	& 303803	&Y	& N \\ 
L1517B	&  045207.2	& 303318	&Y	& N \\
L1512	&  050054.4	& 323900	&Y	& N \\
L1544	&  050114.0	& 250700	&Y	& N \\
L1523	&  050302.7	& 313730	&Y	& N \\ 
L1582A  &  052911.9	&122820  	&Y	&N \\
L1622A	&  055217.1	& 015655	&?	& N \\ 
L1621-1&  055323.0	& 021739	&?	&N  \\ 
CB37	&  055729.9	& 313925	&Y	& N \\ 
L1574	&  060508.9	& 182840	&Y	& N \\ 
L1578-2&  060540.4	& 181200	&Y	& Y \\
CB45	&  060602.0	& 175048	&Y	& N \\
L1633	&  062204.0	& 032353	&Y	& N \\
\enddata
\tablecomments{A question mark in the ``HINSA'' column means that possible absorption is
indicated by a shoulder in the line profile.
The ``EYSO'' column indicates the identification of
young stellar object (or objects) based on IRAS data \citep{lee99}.}
\end{deluxetable}
\normalsize
%
%
%
 \begin{deluxetable}{lccccccccccccc}  
 \rotate  
 \tabletypesize{\scriptsize} 
 \tablewidth{0pt}  
 \setlength{\tabcolsep}{0.04in}  
\tablecolumns{14}  
\tablecaption{Survey of Dark Clouds--Spectral Line Characteristics}  
\tablehead{  
\colhead{Source} & \multicolumn{4}{c}{V$_{lsr}$} & \colhead{p} & \multicolumn{6}{c}{$\Delta V_{nt}$}  
&\colhead{T$_{eq}$} & \colhead{T$^{upper}_x$} \\ 
&\multicolumn{4}{c}{\kms} &  &\multicolumn{6}{c}{\kms}  
&\colhead{K} & \colhead{K}  \\ 
\cline{2-5} \cline{7-12}  \\  
&\colhead{HI} & \colhead{OH} & \colhead{\c18o}  & \colhead{CI}&  &
\colhead{HI} & \colhead{OH} & \colhead{\c18o} & \colhead{\13co} & \colhead{CO} & \colhead{CI} & & \\ }
\startdata 
L1524-1 & 6.33$\pm$0.08\tablenotemark{a} &6.42$\pm$0.05\tablenotemark{a} &6.47$\pm$0.04\tablenotemark{a} & 5.6$\pm$0.4\tablenotemark{a} & 0.8 &0.51$\pm$0.15\tablenotemark{b} &0.75$\pm$0.05\tablenotemark{a} &0.69$\pm$0.06\tablenotemark{a} & 1.0$\pm$0.2\tablenotemark{a} &1.46$\pm$0.07\tablenotemark{a} & 5.3$\pm$0.3\tablenotemark{a} &15.8$\pm$0.5 \tablenotemark{b} &  26 \\ 
L1524-2 &6.39 &6.33 &6.37 & 6.5 &0.80 &0.74 &1.02 &0.75 & 1.0 &1.54 & 2.2&22.1  &  28 \\ 
L1521E &6.57 &6.51 &6.72 & 6.6 &0.81 &0.00 &0.76 &0.51 & 0.7 &0.96 & 1.2& 6.7  &  35 \\ 
B18-2 &6.15 &6.19 &6.20 & 6.8 &0.80 &0.38 &0.54 &0.39 & 0.5 &0.98 & 1.9&13.2  &  29 \\ 
L1536-2 &5.73 &5.79 &5.87 & 6.0 &0.79 &1.00 &0.72 &0.69 & 0.9 &1.21 & 1.6&32.1  &  29 \\ 
TMC2-3 &6.16 &6.23 &6.57 & 6.5 &0.80 &1.53 &1.09 &0.53 & 0.8 &1.16 & 2.5&61.3  &  23 \\ 
L1536-1 &5.58 &5.58 &5.62 & 5.8 &0.80 &0.34 &0.35 &0.39 & 0.7 &1.01 & 1.9&12.5  &  32 \\ 
B18-4 &5.58 &5.73 &5.92 & 6.0 &0.81 &1.34 &0.86 &0.66 & 0.8 &1.13 & 1.9&49.5  &  27 \\ 
L1527A-1 &6.08 &6.17 &6.26 & 6.3 &0.82 &1.28 &0.86 &0.49 & 0.7 &1.22 & 1.3&46.0  &  20 \\ 
L1534 &6.11 &6.15 &6.30 & 6.3 &0.82 &1.42 &0.87 &0.67 & 1.0 &1.38 & 2.5&54.3  &  15 \\ 
L1507-2 &6.36 &6.25 &6.29 & 6.4 &0.87 &0.22 &0.68 &0.50 & 0.7 &1.06 & 4.2&11.1  &  44 \\ 
L1517C &5.36 &5.49 &5.58 & 5.4 &0.90 &0.80 &0.41 &0.50 & 0.5 &0.78 & 3.5&24.2  &  41 \\ 
L1517B-2 &5.86 &5.93 &6.02 & 6.2 &0.90 &0.42 &0.65 &0.61 & 0.7 &0.78 & 1.2&13.9  &  44 \\ 
L1517B &5.78 &5.78 &5.84 & 5.9 &0.90 &0.34 &0.35 &0.35 & 0.4 &0.57 & 1.4&12.6  &  43 \\ 
L1512 &7.00 &7.03 &7.10 & 7.2 &0.90 &0.00 &0.29 &0.27 & 0.4 &0.52 & 1.6& 6.8  &  35 \\ 
L1544 &6.96 &7.10 &7.18 & 7.8 &0.87 &0.49 &0.44 &0.42 & 0.5 &0.65 & 3.8&15.3  &  39 \\ 
L1523 &6.90 &6.96 &7.02 & 7.0 &0.78 &0.00 &0.25 &0.25 & 0.4 &0.41 & 1.9& 6.8  &  28 \\ 
L1582A &10.07 &10.10  &\nodata &10.3  &0.62 & 1.5 &0.55  &\nodata &\nodata &\nodata  & 1.8  &61.1  &  27 \\ 
CB37 &1.26 &1.10 &1.13 & 1.3 &1.00 &1.60 &1.77 &1.45 & 1.7 &2.23 & 2.8&65.7  &  50 \\ 
L1574-b &0.09 &0.13 &0.16 & 0.1 &1.00 &0.89 &1.03 &0.58 & 0.8 &1.46 & 1.7&27.4  &  57 \\ 
L1574-r &3.27 &3.29  &\nodata & 3.5  &1.00 & 1.1 &1.67  &\nodata &\nodata &\nodata  & 3.1  &38.8  &  66 \\ 
L1578-2 &-0.80 &-0.85 &-0.48 &-0.9 &1.00 &1.31 &1.62 &0.74 & 1.0 &1.45 & 2.2&47.6  &  44 \\ 
CB45 &0.76 &0.61 &0.66 & 1.4 &0.95 &0.97 &1.18 &0.73 & 0.9 &1.28 & 1.6&30.4  &  64 \\ 
L1633 &9.65 &9.40 &9.48 & 9.9 &1.00 &1.75 &1.20 &1.85 & 1.9 &2.54 & 3.1&77.1  &  63 \\ 
Average &\nodata&\nodata &\nodata &\nodata&\nodata  & 0.8 &0.83 &0.64 & 0.8 &1.17 & 2.3 &\nodata&\nodata  \\ 
\enddata 
\tablecomments{Columns 2, 3, 4, and 5 give the rest velocities. The non-thermal linewidth, $\Delta V_{nt}$, is defined
by Eq.~\eqref{vnt}. The upper limits to kinetic temperatures, T$_{eq}$ and T$^{upper}_x$, are discussed in section~\ref{temperature}. The column p represents the portion of HI in the background based on a single gaussian disk model (Eq.~\ref{p}). A value of p=1.0 means that the source distance is unknown.} 
\tablenotetext{a}{The one sigma statistical uncertainty is  based on a Gaussian fit. This value is representative of the sample. The uncertainty varies for individual sources due to the difference in signal to noise ratio and in the degree by which a profile deviates from a Gaussian.} 
\tablenotetext{b}{The uncertainty is estimated based on the variations in fitted background temperature when different orders of polynomials are used (Figure~\ref{fig:rta}). The statistical uncertainty is insignificant compared to this.}
\end{deluxetable}

 \begin{deluxetable}{lcccccc}  
 \tablewidth{0pt}  
 \setlength{\tabcolsep}{0.04in}  
\tablecaption{Survey of Dark Clouds--Column Densities}  
\tablehead{  
\colhead{Source} &\colhead{$\tau$(HINLA)} &\colhead{N(HINLA)}  &\colhead{N(OH)} &\colhead{N(\c18o)} &\colhead{N(CI)} &\colhead{[HI]/[H$_2$]} \\ 
 & &\colhead{$10^{18}$ \cm2}  &\colhead{$10^{14}$ \cm2} &\colhead{$10^{15}$ \cm2} &\colhead{$10^{16}$ \cm2} & }
\startdata 
L1524-1 &0.23$\pm$0.04\tablenotemark{a} & 3.9$\pm$0.8  \tablenotemark{a} & 1.4$\pm$0.1  \tablenotemark{b} &1.37$\pm$0.06  \tablenotemark{b} & 4.9$\pm$0.1  \tablenotemark{b} &4.8e-04   \\ 
L1524-2 &0.19 & 3.8   & 1.7   &1.77   & 4.5  &3.6e-04   \\ 
L1521E &0.05 & 0.5   & 0.8   &0.76   & 1.3  &1.2e-04   \\ 
B18-2 &0.15 & 2.3   & 0.7   &0.48   & 1.1  &8.1e-04   \\ 
L1536-2 &0.31 & 7.2   & 1.3   &0.75   & 2.6  &1.6e-03   \\ 
TMC2-3 &0.41 &13.3   & 1.6   &0.98   & 5.0  &2.3e-03   \\ 
L1536-1 &0.18 & 2.6   & 1.2   &0.82   & 2.8  &5.5e-04   \\ 
B18-4 &0.40 &11.7   & 1.8   &1.24   & 4.1  &1.6e-03   \\ 
L1527A-1 &0.83 &23.5   & 1.2   &0.55   & 2.1  &7.3e-03   \\ 
L1534 &1.31 &40.4   & 2.5   &1.69   & 5.8  &4.1e-03   \\ 
L1507-2 &0.02 & 0.3   & 1.1   &0.61   & 2.9  &9.0e-05   \\ 
L1517C &0.15 & 3.2   & 0.6   &0.94   & 2.3  &5.7e-04   \\ 
L1517B-2 &0.11 & 1.7   & 0.8   &0.70   & 1.0  &4.0e-04   \\ 
L1517B &0.11 & 1.6   & 0.7   &0.55   & 2.1  &4.8e-04   \\ 
L1512 &0.06 & 0.6   & 0.6   &0.41   & 2.2  &2.6e-04   \\ 
L1544 &0.29 & 4.7   & 1.7   &1.08   & 5.4  &7.4e-04   \\ 
L1523 &0.13 & 1.4   & 0.9   &0.77   & 1.6  &3.0e-04   \\ 
L1582A &0.21 & 6.8 & 0.4  & \nodata      & 7.0  &\nodata   \\ 
CB37 &0.19 & 6.6   & 1.5   &0.95   & 2.2  &1.2e-03   \\ 
L1574-b &0.22 & 4.8   & 0.7   &0.21   & 1.9  &3.9e-03   \\ 
L1574-r &0.29 & 7.6 & 0.8  & \nodata      & 2.6  &\nodata   \\ 
L1578-2 &0.39 &11.1   & 1.0   &0.43   & 2.2  &4.4e-03   \\ 
CB45 &0.20 & 4.6   & 0.6   &0.26   & 1.7  &2.9e-03   \\ 
L1633 &0.21 & 7.9   & 0.3   &0.38   & 3.6  &3.6e-03   \\ 
Average &0.28 &7.2e+18   &1.1e+14   &8.0e+14   &3.0e+16  &1.5e-03   \\ 
\enddata 
\tablenotetext{a}{The uncertainty is estimated based on the variations in fitted background temperature when different orders of polynomials are used.}
\tablenotetext{b}{These are the one sigma statistical uncertainties representative of the sample.}\end{deluxetable} 

\end{document}